\documentclass[conference,compsoc]{IEEEtran}

\usepackage[camera]{dtrt}

\usepackage{amsmath}
\usepackage{amsthm}
\usepackage{hyperref}
\usepackage{algpseudocode}
\usepackage{booktabs}
\usepackage{boxedminipage}
\usepackage{cite}
\usepackage[ff,mm,sets,adversary,advantage,asymptotics,probability,keys,primitives,operators,lambda,logic,notions]{cryptocode}
\usepackage{enumitem}
\usepackage{listings}
\usepackage{multirow}
\usepackage{subcaption}
\usepackage{threeparttable}
\usepackage{pifont}
\usepackage{xspace}

\definecolor{linkcolor}{rgb}{0.7,0,0}
\hypersetup{colorlinks = true, urlcolor = blue, linkcolor = blue, citecolor = blue}

\renewcommand{\appendixautorefname}{Appx.}
\newcommand{\appendixsubsecref}[1]{\hyperref[#1]{\appendixautorefname~\ref{#1}}}

\newif\iffull
\fulltrue
\iffull
    \newcommand{\FULL}[1]{#1}
    \newcommand{\CONF}[1]{}
\else%
    \newcommand{\FULL}[1]{}
    \newcommand{\CONF}[1]{#1}
\fi

\newif\ifideas
\ideastrue
\ifideas
    \newcommand{\IDEA}[1]{#1}
    \newcommand{\SOLUTION}[1]{}
\else
    \newcommand{\IDEA}[1]{}
    \newcommand{\SOLUTION}[1]{#1}
\fi

\newcounter{finding}
\newcommand{\finding}[2]{
\vspace{6pt}
\noindent
\framebox{
\begin{minipage}[b]{0.95\columnwidth}
\noindent \textbf{Finding \Roman{finding}}: \textit{#1}
\stepcounter{finding}
\end{minipage}
}\vspace{6pt}}

\newcommand{\numberofauctions}{301,479\xspace}

\newcommand{\ultrasound}{ultra sound relay\xspace}

\definecolor{codebg}{rgb}{1,1,1}
\definecolor{codegreen}{rgb}{0,0.5,0}
\definecolor{codeblue}{rgb}{0.13, 0.13, 0.67}
\definecolor{codered}{rgb}{0.75, 0, 0}
\definecolor{codemagenta}{rgb}{0.58, 0, 0.82}
\definecolor{codegray}{rgb}{0.5,0.5,0.5}

\definecolor{myblue}{RGB}{70,130,180}
\definecolor{mydeepblue}{RGB}{65,105,225}
\definecolor{myviolet}{RGB}{97,0,138}
\definecolor{myburgundy}{RGB}{110,10,30}
\definecolor{mygreen}{RGB}{0,105,148}
\definecolor{mygrey}{RGB}{180, 180, 200}
\definecolor{idealfun}{RGB}{165,42,42}
\definecolor{check}{RGB}{11,141,10}
\definecolor{auburn}{rgb}{0.43, 0.21, 0.1}
\definecolor{codepurple}{rgb}{0.58,0,0.82}
\definecolor{codebrown}{rgb}{0.6, 0.4, 0.2}
\definecolor{codeyellow}{HTML}{D19A66}
\colorlet{party}{myburgundy}
\colorlet{protstring}{myviolet}
\colorlet{comment}{mygrey}

\title{Decentralization of Ethereum's Builder Market}

\begin{document}

\author{\IEEEauthorblockN{Sen Yang}
\IEEEauthorblockA{
Yale University\\
sen.yang@yale.edu}
\and
\IEEEauthorblockN{Kartik Nayak}
\IEEEauthorblockA{
Duke University\\
kartik@cs.duke.edu}
\and
\IEEEauthorblockN{Fan Zhang}
\IEEEauthorblockA{
    Yale University\\
    f.zhang@yale.edu
}}

\maketitle

\begin{abstract}
Blockchains protect an ecosystem worth more than \$500bn with strong security properties derived from the principle of {\em decentralization}. Is today's blockchain decentralized? In this paper, we empirically studied one of the {\em least decentralized} parts of Ethereum, its builder market.

The builder market was introduced to fairly distribute Maximal Extractable Value (MEV) among validators and avoid validator centralization.
As of the time of writing, {\em two builders} produced more than 85\% of blocks in Ethereum, creating a concerning centralization factor. However, a common belief is that such centralization ``is okay,'' arguing that builder centralization will not lead to validator centralization.
In this empirical study, we quantify the significant proposer losses within the centralized builder market and challenge the belief that this is acceptable.

The significant proposer losses,
\done%
if left uncontrolled, could undermine the goal of PBS. Moreover, MEV mitigation solutions slated for adoption are affected too because they rely on the builder market as an ``MEV oracle,'' which is made inaccurate by centralization.
Our investigation reveals the incentive issue within the current MEV supply chain and its implications for builder centralization and proposer losses.
Finally, we analyze why the proposed mitigation cannot work and highlight two properties essential for effective solutions.

\end{abstract}

\section{Introduction}

At the core of any decentralized blockchain, a key assumption underpinning their strong security properties (integrity, availability, censorship resistance, incentive compatibility, etc.) is ``decentralization'', that the system is run by a large number of independent participants~\cite{nakamoto2008bitcoin,eyal2018majority,schwarz2022three,wahrstatter2023blockchain,bahrani2024centralization,roughgarden2020transaction}.
It is community consensus that one of the greatest threats to decentralization is Maximal Extractable Value (MEV)~\cite{daian2020flash},
profits that protocol participants, known as validators, can reap by manipulating the ordering of transactions.
MEV can lead to validator centralization because it disproportionally benefits well-resourced validators compared to ``regular'' ones, as extracting MEV requires significant capital, intelligence, and computational resources, which regular validators may not have~\cite{bahrani2024centralization, EthereumPBS2023, pai2023structural,buterin2021pbs,buterin2021endgame}.

To distribute MEV fairly among validators and avoid validator centralization, Ethereum adopted a solution called Proposer-Builder Separation (PBS)~\cite{EthereumPBS2023}.
A proposer is the validator chosen to add a new block to the blockchain.
In the current implementation called MEV-Boost~\cite{flashbots2022mevboost}, %
the proposer runs an auction among builders (anyone can be a builder) to solicit the best block.
Ideally, builders compete with each other, and the proposer collects a significant portion of MEV in auction revenue without doing much work, eliminating the advantage of sophisticated validators over regular ones. 

\begin{figure}[tbp]
    \centering
    \includegraphics[width=\columnwidth]{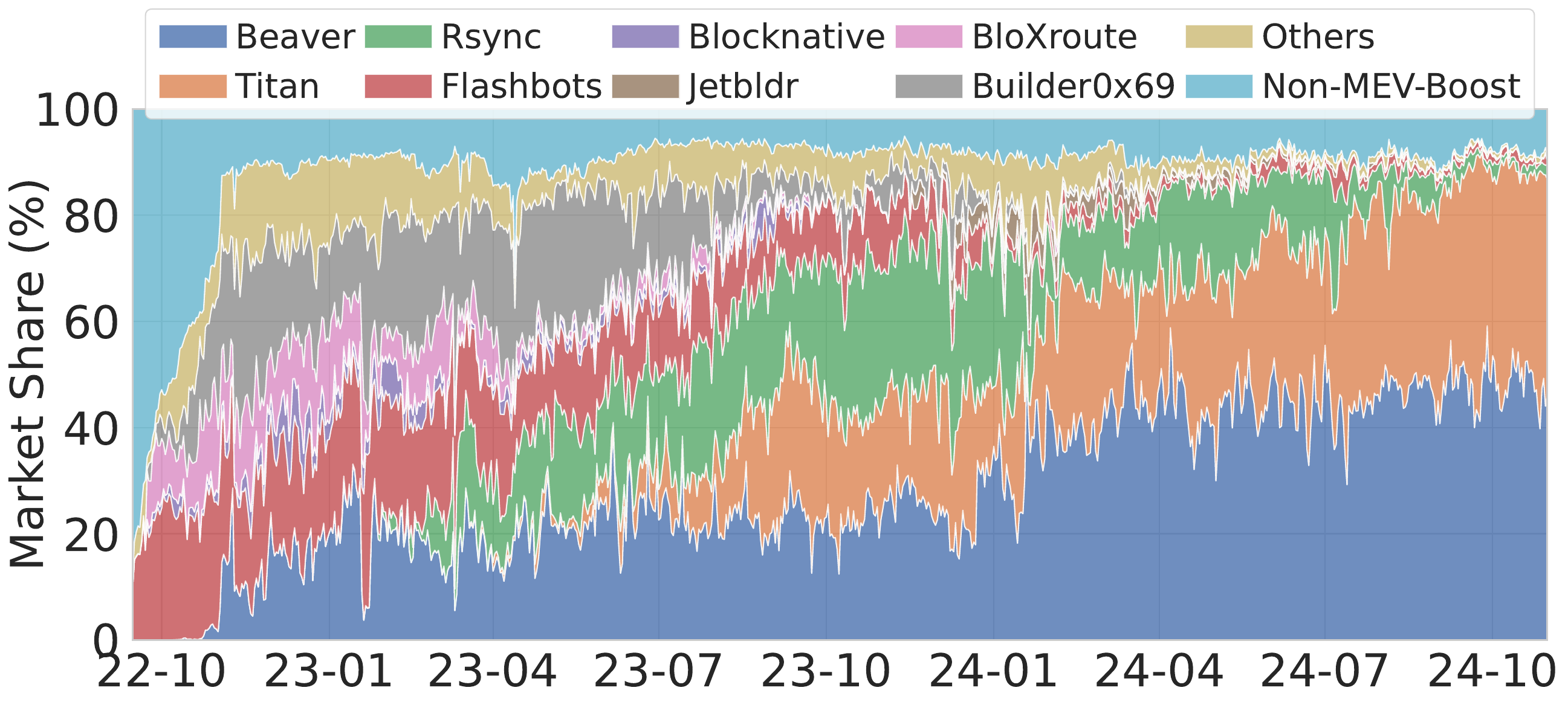}
    \caption{Market share of builders (the source of the data is introduced in~\autoref{sec:dataset}). The builder market is one of the least decentralized parts of Ethereum!}
    \label{fig:centralized-builder-market}
\end{figure}

\done%

\emph{Is the builder market working as intended?} The answer is rather unclear because the builder market is highly centralized.
As shown in~\autoref{fig:centralized-builder-market}, two builders won over 90\% of auctions.
The builder centralization creates immediate security problems like censorship~\cite{wahrstatter2023blockchain}.
However, the community still believes that ``it is OK to centralize block building''~\cite{EthereumPBS2023}, arguing that block centralization will not lead to validator centralization. 
{\bf We challenge this belief through an empirical study of the past MEV-Boost auctions.}
As will be shown in detail, proposers have incurred significant losses in a centralized builder market and, if left uncontrolled, could lead to undesired consequences including undermining PBS.

\subsection{Proposer Loss in the MEV-Boost Auctions}

\parhead{Proposer loss and its consequences}
Intuitively, a proposer's loss is the difference between the MEV available in a block and the auction revenue.
Proposers may incur losses in a centralized builder market in two ways. First, auctions may not be competitive enough due to, e.g., collusion or other limitations of the auction mechanism.
Second, even if builders compete with each other, if a single builder has a unique advantage, it will not be incentivized to bid on all of the MEV it can extract. Instead, it only needs to outbid the second-best bidder slightly to win.
We say that the proposer incurs a {\em loss from uncompetitiveness} in the first case and a {\em loss from inequality} in the second case.

If proposers suffer significant losses, they would be incentivized to extract MEV {\em by themselves} (instead of using the builder market). This is undesirable because it re-introduces the dynamics that PBS wants to avoid: sophisticated validators would have an advantage over small ones in extracting MEV.
Moreover, several MEV mitigation solutions rely on the MEV-Boost auctions as an ``oracle'' to know how much MEV has been extracted from the system (e.g., MEV-burn~\cite{mevburn2024} proposed to remove that amount from circulation, effectively redistributing them to all holders of ETH through deflation.) 
The proposer's loss translates to inaccuracy in the MEV oracle.

\done%

Given the undesired consequences of proposer losses, it is important to understand the extent of losses incurred by proposers in the current builder market. Next, we introduce the auction mechanism and present how we quantify proposer losses using auction data.

\parhead{Modeling MEV-Boost auctions}
MEV-Boost auction adopts an open-bid, ascending price auction akin to an English auction~\cite{besanko2009economics}, with a 12-second deadline.
For each auction instance, builders continuously submit bids in the form of $(B,BV)$ specifying a block $B$ and a {\em bid value} (BV) the builder is willing to pay for $B$ to be added to the blockchain.
The revenue a bidder gains from $B$ is called the \textit{true value} ($TV(B)$), which we estimate from transactions in $B$.
At the end of an auction, the block $B$ with the highest bid is added to the blockchain, the builder pays $BV$ to the proposer and earns $TV$ when $B$ is executed (i.e., the builder's profit is $TV-BV$).

\begin{figure}
\includegraphics[width=.9\columnwidth]{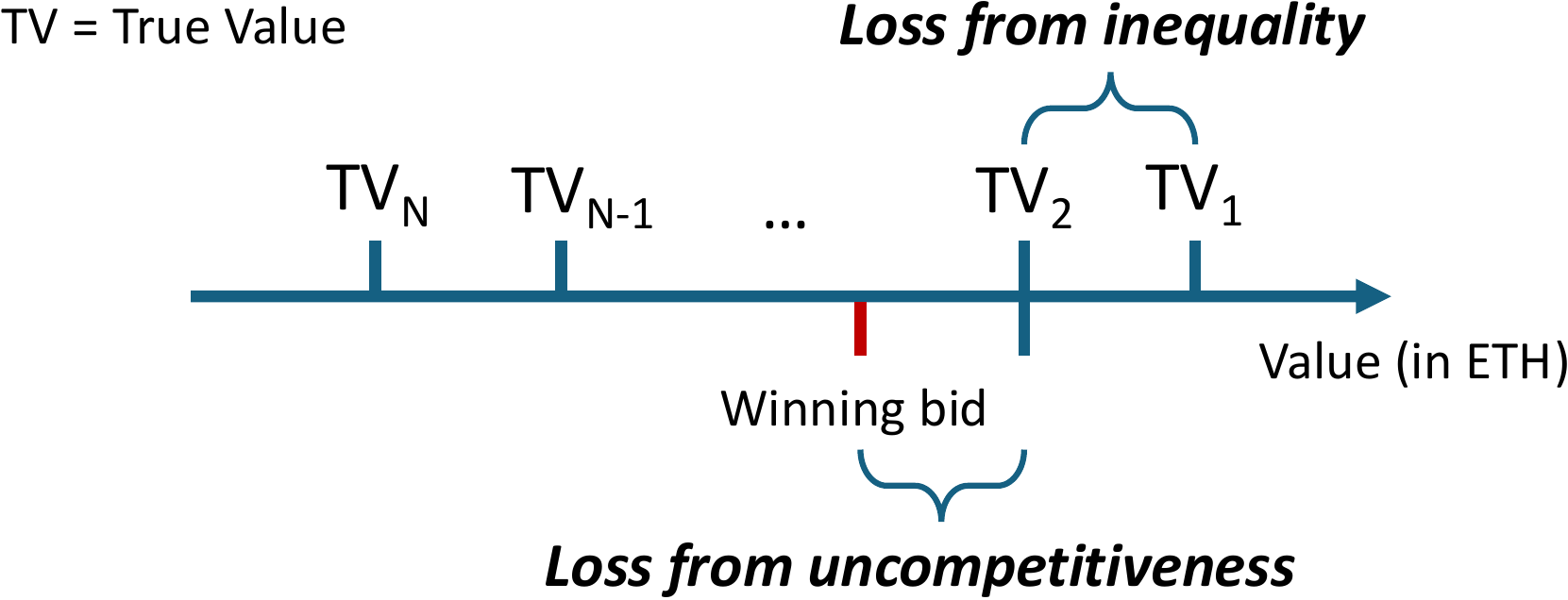}
\caption{Definition of proposer losses.}
\label{fig:losses}
\end{figure}

The two types of losses defined above can be computed from auction bids, as illustrated in~\autoref{fig:losses}.
Specifically, suppose the true values of the bids from a given auction instance are $TV_1 > TV_2 \ge \cdots \ge TV_N$.
The loss from inequality is $TV_2-TV_1$ because a competitive auction can, at most, generate revenue of $TV_2$.
The loss from uncompetitiveness is $\max(TV_2 - BV, 0)$; if $TV_2 - BV$ is negative, the auction is considered competitive without this loss.

\parhead{Auction data}
A practical challenge in quantifying the proposer loss in past MEV-Boost auctions is the lack of data. Only the winning bids are recorded in the blockchain, but to compute the losses, we also need losing bids.
The key ingredient that enabled our study is a large-scale auction dataset we curated since 2022: we collected approximately 12 billion partial bids\footnote{Partial and full bids are to be defined in~\autoref{sec:data-collection}} of auctions from September 2022 to October 2024; in addition, in collaboration with \ultrasound~\cite{UltraSoundRelay}, we obtained full bids for a total of \numberofauctions auctions from April 2023 to December 2023. From the auction dataset, we can compute the true value of all bids.

\parhead{Results} 
91.8\% of the MEV-Boost auctions we analyzed were competitive. However, auctions tend to be less competitive when the MEV of a slot increases.
The loss from uncompetitiveness turns out rather insignificant, ranging from 0.5\% to 1.7\% during the periods of our study.

To understand the variations in builders' abilities to extract values, we statistically compared the true values across different builders in the same auction.
We used several metrics, including the quartile coefficient of dispersion (QCD)~\cite{bonett2006confidence}, to quantify the inequality of true values across builders.
We found significant differences in the block-building capacities of top, middle, and tail builders. The inequality worsens as the MEV of a slot increases, likely because significant MEV opportunities are shared with only a limited number of builders.

Inequality of block-building capacities has led to a proposer loss ranging from 5.6\% to 11.5\%. I.e., had there been no inequality, the proposer would have earned roughly 10\% more. This indicates that inequality is the primary driver of the proposer's losses in the MEV-Boost auctions.

\subsection{Private Order Flows}
Given that proposers suffer significant losses due to the builders' block-building capacity inequality, we now investigate the underlying causes of this inequality.

In the current MEV ecosystem (c.f.~\autoref{sec:mev-supply-chain}), builders receive {\em order flows} from order flow {\em providers} and compute the optimal ordering of received order flows (see~\cite{flashbots2024rbuilder} for a reference builder algorithm).
Thus, a builder's block-building capacity is determined by the ``quality'' of the order flows it receives.
At the same time, not all order flows are accessible to all builders. The difference in the builder's ability to access {\em important} private order flows can drive the inequality in block-building capacity. 

\done%
Not every provider is equally important, and some providers that generate profitable order flows can significantly impact builders.
Thus, we propose a metric to identify important providers and investigate their implications.

\parhead{Identifying pivotal private order flow providers.}

To formally capture the intuition of ``important providers'', we defined \textit{pivotal level} to quantify the sustained impact of a given provider.
A provider $P$ is {\em pivotal} in an auction $A$ if the winner of $A$ would have lost, had transactions from $P$ not been available. In other words, pivotal providers are necessary for the winner to win.
The pivotal level of a provider $P$ is the percentage of MEV-Boost auctions in which $P$ is pivotal. 
Other metrics, calculated based on each provider's contribution to the builders' blocks~\cite{lu2023illuminating, OrderflowArt2024}, may include temporarily influential providers, whereas this metric identifies sustainedly influential providers, resulting in fewer false positives.
\done%

Using the auction dataset, we identified five pivotal providers whose pivotal level exceeds 50\% over a period longer than two weeks.
That is, if a builder cannot access the flows from any of these providers, it will lose the majority of the auctions during our study period.

\done%
Order flows from these pivotal providers have two levels of accessibility, which we introduce below.

\parhead{ Accessibility by multiple builders} Among the top 5 pivotal providers, MEV-Share and MEV Blocker send their order flows to all builders they consider reputable.
They do not cause inequality loss since the competition among these builders allows proposers to capture this value.

The reputability of a builder is assessed per their specific rules.
While top builders usually pass the assessments, it is harder for new builders to enter the market.

\parhead{Exclusive pivotal providers}
A more worrisome finding is that some providers send order flows {\em almost exclusively} to their preferred builders.
Besides the well-known exclusive relationships (known as ``integration''~\cite{gupta2023centralizing}), we also identified that three of the top five pivotal providers, previously not known to have integration, began integrating with top builders in 2024.
\done%
Our findings show that the integration of these three pivotal providers accounted for significant proposer losses in 2024.

Integration poses a market entry barrier as new builders likely lack integrated providers. 
Moreover, it offers a unique competitive edge and will likely cause dominance. 
Finally, there is a strong incentive to integrate: compared to sending valuable order flows to multiple builders, sending them to a single builder enables this builder to win the auction with higher profit; the provider and the builder can share the profit, which means integration improves their total utility. 

Integration is detrimental to proposers, as it will reduce their profits. The strong incentive for integration makes it increasingly difficult to address builder centralization.

\subsection{Mitigation}

Our study reveals trust concerns and incentive issues within the MEV supply chain. They cannot be resolved by the proposed alternative PBS designs~\cite{epbs2023,mevburn2024,executiontickets2023,monnot2024proposers}, as these designs may retain the same underlying supply chain.

To fully address these problems, we identify two essential properties: a security guarantee protecting providers from malicious builder attacks and a game-theoretical guarantee disincentivizing provider integration. While solving these problems remains a grand interdisciplinary challenge requiring further study, identifying these essential properties lays a foundation for future solutions.

\done%

\subsection*{Summary}
To the best of our knowledge, we are the first to use the rich auction data to analyze the decentralization of Ethereum's builder market, while most previous works are limited to on-chain data. \done%
As a result, they cannot identify the primary cause of proposer losses or accurately identify important providers.
We summarize our contributions below.

\begin{itemize}[leftmargin=*]
    \item We found that proposers incur significant losses in the centralized builder market, which could lead to undesirable consequences, including validator centralization and inaccurate MEV oracles. We challenge the community's belief that ``builder centralization is okay".
    \item 
    We attribute proposer losses to two factors: auctions are not competitive, builders' block-building capacities are unequal, and inequality is the primary cause.
    \item We identify a previously unrecognized integration of top providers,  which causes inequality in block-building capacity and significant losses for proposers.
    \item Our dataset of the true values of \numberofauctions MEV-Boost auctions from April to December 2023 may be of independent interest.
Our dataset of approximately 12 billion partial bids since 2022 is the largest scale known to us. We have open-sourced the code and datasets at \cite{decentralizationBuilderMarketRepo} and \cite{decentralizationBuilderMarketData}.
\item Besides contributing to the literature on MEV mitigation, our work also contributes to the Economics literature on the analysis of MEV-Boost auctions. 

The findings in this paper can serve as empirical evidence for future theoretical interrogation.
\end{itemize}

\begin{figure*}[!tbp]
    \centering
    \includegraphics[width=0.95\textwidth]{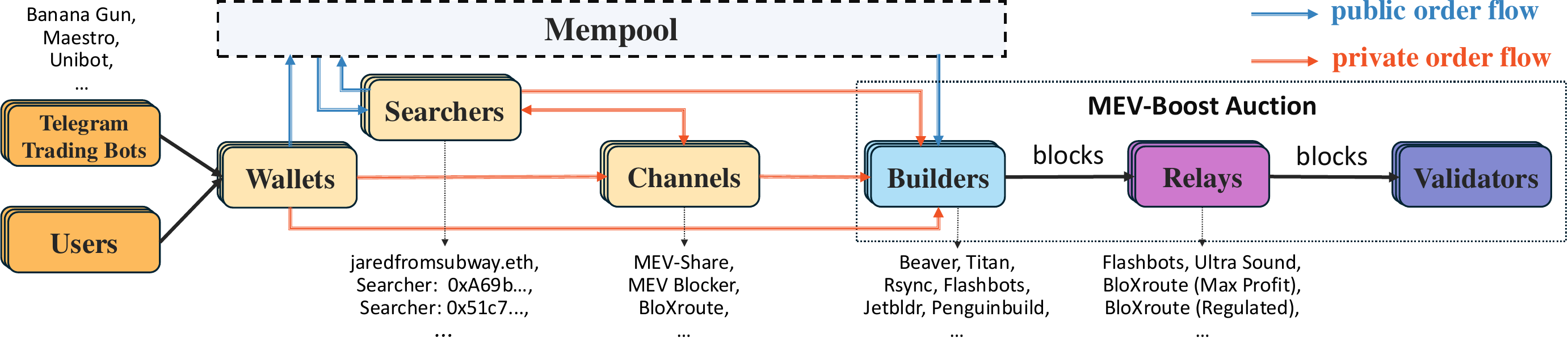}
    \caption{Illustration of the MEV supply chain in the PBS scheme.}
    \label{fig:background}
\end{figure*}

\section{Background}
\label{sec:background}

\subsection{MEV-Boost Auctions}
\label{sec:pbs}

\parhead{MEV}
Miner/Maximal Extractable Value (MEV)~\cite{daian2020flash} refers to the profit that privileged players  (e.g., validators) can earn by including, excluding, and reordering the transactions. MEV poses a threat to decentralization, as explained in the introduction. 
There are two schools of thought on mitigating the negative impacts of MEV. The first one is referred to as MEV democratization~\cite{flashbots2022mev}, the main idea of which is to facilitate MEV extraction and level the playground. MEV-Boost~\cite{flashbots2022mevboost}, to be introduced below, is the primary example. Another approach is to enforce certain ordering policies (e.g., FIFO) to prevent order manipulation, such as~\cite{kelkar2020order, kelkar2023themis}. 
We focus on MEV-Boost because it is widely used.

\parhead{PoS Ethereum}
Before introducing background on MEV-Boost auctions, we first cover the relevant concepts in Ethereum's Proof of Stake-based consensus protocol called Gasper~\cite{buterin2020combining}. 
Participants of Gasper are called \textit{validators}. Each validator must deposit 32 ETH as collateral~\cite{ethereum2024staking}.
Time is divided into 12-second intervals called {\em slots}. This results in 7,200 slots in Ethereum per day. During each slot, a single validator, referred to as the \textit{proposer}, proposes a block, while others vote for the chain head in their local view.

\parhead{Proposer-builder separation (PBS) and MEV-Boost}
PBS can be conceptually viewed as a sidecar to the core consensus protocol (although some proposals are more tightly coupled with consensus) to allow the proposer to outsource the job of building blocks to {\em builders}. 

MEV-Boost~\cite{flashbots2022mevboost} is the current incarnation of the PBS idea. About 90\% of all blocks on Ethereum are built by MEV-Boost~\cite{mevboost2024}. The basic idea is to auction off the right to propose a block to the builder who bids the highest in an open-bid ascending price auction~\cite{english_auction_wiki}.
Because builders and proposers are mutually untrusted, the auction is facilitated by a (trusted) third party called a relay~\cite{flashbots2022mevboostrelay}.
Here are the high-level steps (we refer readers to \cite{flashbots2022mevboost} for details):

\begin{itemize}[leftmargin=*]
\item Builders submit bids to relays throughout the slot. For this paper, we denote a \textit{full bid} with $(B, BV, PK_\text{builder})$ that consists of a block $B$, a bid value $BV$ to be paid to the proposed if the block is proposed, and the builder's public key. 
Bids are signed by the builder, but we omit the signature for clarity.
Relays verify that $B$ indeed pays the proposer at least $BV$.
\item Relays expose a public API that publishes {\em partial bids} in the form of $(\mathsf{Hash}(B), BV, metadata)$. $metadata$ includes parent hash, gas limit, gas used, etc.
Partial bids do not include block content.
\item The proposer selects a block hash $h_B$, typically the one with the highest $BV$, and sends the relay a signature on $h_B$. This signature is a promise to propose $B$ in the consensus protocol (if it proposes another block, this signature is evidence of equivocation and will cause the proposer to be penalized~\cite{ethereum2025consensus}.)  After verifying the signature, the relay returns $B$ to the proposer.
\end{itemize}

Two additional remarks. First, relays are trusted entities, but anyone can run a relay (currently ten~\cite{ethstaker2024mevrelaylist}), and builders and proposers can connect to the relays they trust. 
Second, winning bids are recorded on-chain, and losing bids are discarded after the auction.
We collected and archived the historical bids over time to enable this study, as we will detail in~\autoref{sec:dataset}.

\subsection{MEV Supply Chain}
\label{sec:mev-supply-chain}

The ecosystem around discovering and extracting MEV forms an MEV supply chain, as depicted in \autoref{fig:background}. 
Builders, relays, and proposers (validators) have been introduced above. Now we introduce the ``upstream'' of builders. %

\parhead{Wallet, users, trading bots}

Users use wallets to interact with blockchains.
A user communicates her intent (e.g., transferring cryptocurrency) to the wallet, which generates the transaction to fulfill the user's intent and sends it to the network. 
MEV may be generated during this process, e.g., when the user transaction creates an arbitrage opportunity.
While the wallet is typically controlled by the users, it can also be managed by a third party, such as \textit{Telegram trading bots}~\cite{arkham2024telegram}. These bots are automated programs integrated with the Telegram app~\cite{Telegram2024}, automatically conducting cryptocurrency trades on behalf of users. 
Users' transactions eventually reach builders or proposers, either through the peer-to-peer network (the mempool) or through third-party services called private channels.

\parhead{Mempool} Mempool is a temporary storage of pending transactions before they are added to the blockchain. All transactions stored in the mempool are publicly visible.

\parhead{Channel} A private channel, or channel for short, connects users directly with builders, bypassing the mempool. Channels receive user transactions and forward them to the user-specified builders, offering privacy and atomicity guarantees. 
E.g., a user can send a bundle (list) of transactions and request them not to be unbundled. Examples include MEV-Share~\cite{FlashbotsMEVShare}, MEV Blocker~\cite{MEVBlocker}, BloXroute~\cite{bloXrouteBackrunMe}, etc.

\parhead{Searcher} Searchers are a special type of users. They run (typically proprietary) algorithms to construct profitable transaction sequences (bundles) based on the pending transactions and the blockchain state.
Same as regular users, searchers may submit transactions to mempool or channels. 
In addition, most builders have public APIs that allow searchers to submit bundles directly.
Some searchers may prefer channels for the privacy and atomicity benefits.
A searcher is said to {\em integrate} with a builder if it sends almost all of its orders to that builder exclusively. 
For example, SCP (addresses are listed in~\autoref{tab:searcher-list}) is considered an integrated searcher of the builder Beaver~\cite{heimbach2024non}.

Some channels, such as MEV-Share and MEV Blocker, allow searchers to read pending transactions in the channel and extract MEV from them, provided that such extraction should not harm users (e.g., only backrunning is allowed). 

\parhead{Private order flow} Order flow refers to a stream of transactions that may carry extractable value and can be categorized into two categories: public order flow from the public mempool, and private order flow, sent directly by users or searchers without being broadcast to the mempool.
The entities that can provide private order flows to the builders are referred to as \textit{private order flow providers}, which include searchers, channels, users, and Telegram trading bots as shown in~\autoref{fig:background}.

\subsection{Model and Assumptions.}
\label{sec:model}

\parhead{Accounts and states} We follow the model of~\cite{babel2023clockwork, babel2023lanturn}. We use $\mathcal{A}$ to denote the space of all possible accounts. For example, in Ethereum, both user-owned and contract-owned accounts are represented by 160-bit identifiers. For $a \in \mathcal{A}$, $\text{balance}(a)$ denotes the balance of all tokens held in account $a$, and $\text{balance}(a)[\mathbf{T}]$ specifies the balance of token $\mathbf{T}$. For simplicity, $\text{balance}(a)[0]$ is used to denote the balance of the primary token (e.g., ETH in Ethereum). 

\parhead{Transaction and block} A transaction $tx$ is a string whose execution changes the system's state. 
A block is an ordered sequence of transactions. Executing transactions in a block \(B\) transforms an initial state \(s\) into a new state \(s'\), represented by \(s' = \text{action}(B)(s)\).

\parhead{Extracted value (EV)} The extracted value of a block \(B\) in state \(s\) for a participant \(p\) is the changes of \(p\)'s balances after executing $B$. Assuming \(p\) controls a set of accounts \(A_p\), the extracted value of \(B\) for $p$ is:
\begin{equation}
EV_p(B) = \sum_{a \in A_p} \text{balance}_{s'}(a)[0] - \text{balance}_s(a)[0]
\label{eq:EV}
\end{equation}
where \(s' = action(B)(s)\).

\parhead{Builder's income}

We assume a builder $p$'s income from building a block $B$ is $EV_p(B)$, or $EV(B)$ for short since the builder identity is usually encoded in the block. 
We compute EV following a similar approach to that in MEV analysis  tools~\cite{eigenphi2023, flashbots_mev_inspect_py} and previous works~\cite{heimbach2023ethereum, soispoke2023empirical}.
Specifically, $EV(B)$ can be computed as the sum of the transaction priority fees~\cite{EthereumGas} and direct transfers~\cite{flashbots_coinbase_transfer} to the builder's addresses, minus the payment to the proposer and refund to other addresses.
We also refer to the builder's income from a block (or a set of transactions in a block) as the {\em value} of the block (or the set of transactions).

Note that $EV(B)$ is an estimation of the builder's actual income from $B$ because the builder may own accounts out of $A_p$ in~\autoref{eq:EV}, or even receive payments off-chain.
We discuss the implications in~\autoref{sec:discussion}.

\section{Dataset}
\label{sec:dataset}

We created two auction datasets to study the decentralization of the builder market: a dataset of {\em partial bids} from September 2022 to October 2024, and a dataset of {\em full bids} from April to December 2023. The full bids are obtained from the \ultrasound~\cite{UltraSoundRelay}.
Recall from~\autoref{sec:pbs} that a bid consists of a block $B$ and a bid value $BV$, while a partial bid consists of $(\mathsf{Hash}(B), BV, PK_\text{builder},metadata)$, where $metadata$ includes parent hash, gas limit, gas used, etc.
To support analysis, we also collected on-chain data, including blocks and transactions on Ethereum, as well as miscellaneous data, which will be detailed below. 
\autoref{tab:dataset} summarizes the datasets.

\begin{table}[!tbp]
\centering
\small
\resizebox{\columnwidth}{!}{
\begin{threeparttable}
\caption{Overview of collected datasets.}
\label{tab:dataset}
\begin{tabular}{l|l|l|p{1.5cm}}
\toprule
\textbf{Dataset} & \textbf{Content} & \textbf{\# of entries} & \textbf{Source}          \\
\midrule
\multirow{2}{*}{On-chain} &  blocks        &   5,551,675     &  \multirow{2}{*}{Reth node} \\
~ &  transactions  &   852,482,466    &  ~ \\
\hline
\multirow{3}{*}{Auction}   
     & blocks produced by MEV-Boost   & 4,909,683  &  relay API  \\
     & partial bids   & 11,966,097,921 & relay API \\
     & full bids from ultra sound    &  326,574,676 & ultra sound \\
\hline
\multirow{5}{*}{Misc.} & builder public keys & 273 & 
\multirow{5}{*}{see~\autoref{sec:data-collection}} \\
        & builder addresses & 102 & \\
        & searcher addresses & 2,172 &  \\
         & private transactions & 112,218,755 &   \\
         & transaction sources & 74,922,499 &   \\
\bottomrule
\end{tabular}
\end{threeparttable}
}
\end{table}

\subsection{Data Collection and Validation}
\label{sec:data-collection}
\label{sec:data-validation}

\parhead{On-chain data} We run an Ethereum execution client Reth~\cite{paradigmxyz2024reth} along with a consensus client Lighthouse~\cite{sigp2024lighthouse} to collect blocks and transactions from September 15, 2022, to October 31, 2024. The on-chain dataset includes 5,551,675 blocks and 852,482,466 transactions.

\parhead{Data from relay APIs}
We connected to all 13 known relays\footnote{Three relays were inactive during the period of our study.} and collected the following information from their public APIs: 1) the winning bid including the final block sent to the proposer; 
and 2) partial bids for blocks that did not win the auction. Recall that partial bids do not include the block body. We collected 4,909,683 winning bids and 11,966,097,921 partial bids from September 15, 2022, to October 31, 2024.
We use winning bids to identify blocks produced through MEV-Boost and partial bids to quantify the importance of private order flow providers in~\autoref{sec:openness}.

To validate the completeness of the data collected from relay APIs, we compared the winning bids we collected with three public datasets~\cite{eden_public_data_overview,mevboost2024,dataalways2023mevboostdata}.
Our dataset includes 720 more winning bids than other datasets. 
Upon comparison with on-chain blocks, we find they are all included on Ethereum, indicating a high level of completeness.

\parhead{Full bids from \ultrasound}
From \ultrasound~\cite{UltraSoundRelay}, we obtained the full bids in  \numberofauctions historical auctions that took place between April and December 2023 (\ultrasound stopped collecting bids at the end of 2023). Since the data volume is large (we record roughly 200 GB of data per day), our dataset samples one week from each month: April 9-15 and May 1-7 to December 1-7, 2023. The dataset comprises 326,574,676 full bids (block, bid value, and builder public key). We use full bids in~\autoref{sec:competition} to quantify the competitiveness and efficiency of MEV-Boost auctions, as well as the proposers’ losses.

The original data from relays includes bids submitted after the deadline due to network latency. We removed them as we are only interested in the bids before the auction ends. 
On April 18, 2023, MEV-Boost introduced bid cancellation that allows a bidder to cancel a previous bid by submitting a lower one. However, proposers can query bids before they are canceled and store them locally~\cite{ethresear_ch_bid_cancellations}, effectively bypassing cancellation. 
Therefore we do not consider cancellation.

Not all builders connect to the \ultrasound, so we need to validate that the full bids from \ultrasound cover a significant subset of all builders.
We compute the percentage of the builders that submit to \ultrasound over all known builders. 
The set of all known builders is a union of all builders that appeared in our datasets, including winning bids, partial bids, and bids from the \ultrasound. 
As shown in~\autoref{fig:bids-validation}, bids from the \ultrasound covered more than 80\% of the builders in over 75\% of the MEV-Boost auctions from April 2023 to December 2023.
Moreover, the top-5 builders by market share in this period (Flashbots, Beaver, Rsync, Builder0x69, and Titan) almost always submitted to \ultrasound.

\parhead{Miscellaneous data}

In practice, builders are known by their common names (e.g., Flashbots builder), but a builder may use multiple public keys to sign bids and use multiple addresses to receive and make payments.
We collect builders' public keys to attribute bids to builders. We collect builder's addresses to calculate the true value of bids.
We use private transaction hashes to identify private order flows and use the searchers' addresses and sources of transactions to identify the private order flow providers in~\autoref{sec:openness}.

\begin{itemize}[leftmargin=*]
\item Builder public keys: We collect public keys controlled by each builder combining two data sources: the public keys in builders' official documents; and the distinct marks in their blocks' ``extra field'' (see~\autoref{appx:data-collection} for details). 

To validate the completeness of our dataset, we manually cross-check it with public datasets~\cite{RatedNetwork2023BuilderPubkeys, edennetwork2024pubkeys} and related works~\cite{heimbach2023ethereum,heimbach2024non}, and the results show that our dataset includes 154 more public keys, so our dataset is the most complete one so far. 
\item Builder addresses: We collect the builder addresses from the blocks' last transactions combined with the public dataset~\cite{EtherscanMevBuilder2024}. Similarly, to validate their completeness, we manually cross-check them with related works~\cite{heimbach2023ethereum,heimbach2024non}, and the results show that our dataset includes 86 more builder addresses, which is the most complete so far.
\item Searcher addresses: We obtained known searcher addresses from Etherscan~\cite{etherscan2024mevbot}, libMEV~\cite{LibMEVLeaderboard2024}, and related works~\cite{heimbach2024non,oz2024wins}.
\item Private transaction hashes: We use data from Mempool Guru~\cite{yang2022sok} to label transactions that bypassed the public mempool as {\em private}. 
We believe these labels have already been cross-validated, as the private transactions were not collected by any of the seven distributed nodes.
\item Transaction sources: We are interested in where a transaction originates from and which part of the MEV supply chain it has traversed through.  We identified transaction sources using ground truth data (as described in~\autoref{appx:data-collection}).

\end{itemize}

\section{Proposer Loss in the MEV-Boost Auctions}
\label{sec:competition}

In this section, we focus on {\em the status quo} and study the implications of a centralized builder market as we have today.
In particular, we are interested in the losses incurred by proposers in the MEV-Boost auctions, as significant losses could lead to undesirable consequences, including validator centralization and inaccurate MEV oracles.

In principle, builder centralization does not imply a significant proposer loss, as optimal outcomes require only a few non-colluding builders with similar block-building capacities to maintain competitive auctions.
To evaluate these two conditions in the centralized builder market, we quantify proposers' losses in historical MEV-Boost auctions by considering two factors: the competitiveness of the auctions and the inequality in block-building capacity.

\subsection{Competitiveness of MEV-Boost Auctions}
\label{sec:quantifying-auction}
\done%

Ideally, revenue from MEV-Boost auctions should be near optimal due to competition. However, whether real-world auctions induce such near-perfect competition is complicated by several confounding factors: builders may not be able to respond to bids due to network latency, or they may collude implicitly or explicitly.
In this section, we quantify the competitiveness of the MEV-Boost auctions using what we call competitive index (CI).
We also look at the allocation efficiency as it evaluates whether the current mechanism induces the desired outcomes.

\parhead{True value} In the context of auctions generally, the {\em true value} of the item being auctioned refers to its actual worth to the bidder.
Recall that in MEV-Boost auctions (\autoref{sec:background}), a builder submits a block $B$ (a list of transactions) and a {\em bid value} $BV$ that the builder is willing to pay to the proposer if $B$ is proposed.
The {true value} of $B$, denoted $TV(B)$, is typically higher than $BV$. 
One can think of $TV$ as the builder's revenue and $BV$ as the cost. The difference $EV:=TV-BV$ is the profit pocketed by the builder, which we call {\em extracted value} (EV)---i.e., $EV(B)$ is the difference of the builder's ETH balance after executing transactions in $B$. For example, suppose executing $B$ yields a net profit of $0.1$ ETH after paying $0.9$ ETH in bid value in the auction, then the true value (the revenue) is $1.0$ ETH.

Thanks to the \textit{full bids dataset}, we can compute not only the transactions that appear in on-chain blocks but also those included in other bids submitted in the auction.
To compute $TV(B)$ of a given block $B$, we follow the standard approach in MEV analysis tools~\cite{eigenphi2023, flashbots_mev_inspect_py} and previous works~\cite{heimbach2023ethereum, soispoke2023empirical}, though we built our own analysis toolchain and optimized its performance to handle the large volume of historical auction data (326.6 million full bids).
We compute $EV(B)$ using a modified Ethereum execution client (Reth~\cite{paradigmxyz2024reth}) to execute transactions in $B$ and compute the net ETH balance changes of the builder's addresses. 
Most builders' addresses are well-known partly because they need to build up a reputation. We cross-validated it with other sources to ensure that our dataset was the most complete one to date (see~\autoref{sec:data-validation} for details.)
Then, with bid value $BV(B)$ from the auction dataset, the true value $TV(B)$ is the sum of $EV(B)$ and $BV(B)$. 

In every slot $s$, builders can submit multiple bids throughout the duration of the slot. We use $BV(p,s)$ to denote the highest bid value from builder $p$ in slot $s$, and $TV(p,s)$ the corresponding true value.

\parhead{Competitive index}
In a given slot $s$, suppose builders $p_1,\cdots,p_n$ are ordered by their true value from high to low, i.e., \(TV(p_1, s) \geq \dots \geq TV(p_n, s)\), $BV_w(s)$ is the bid value of the winner and $TV_w(s)$ is the corresponding true value, the competitive index of slot $s$, \(CI(s)\) is defined as:
\[
    CI(s) = \frac{BV_w(s)-TV(p_2, s)}{TV(p_2, s)} \times 100\%.
\]

Namely, $CI(s)$ measures the relative difference between the winning bid and the second-highest true value. We define an auction as competitive if $CI(s) \geq 0$, meaning the winning bid is at least as high as the second-highest true value; if $CI(s) < 0$, the auction is uncompetitive.

\parhead{Efficient index}
Same as above, the efficient index, \(EI(s)\), is defined as:
\[
    EI(s) = \frac{TV_w(s)-TV(p_2, s)}{TV(p_2, s)} \times 100\%.
\]
$EI(s) > 0$ indicates that the winner has the highest true value, whereas $EI(s) \leq 0$ implies that the bidder with the highest true value lost the auction.

\begin{figure}
    \centering
    \includegraphics[width=0.95\columnwidth]{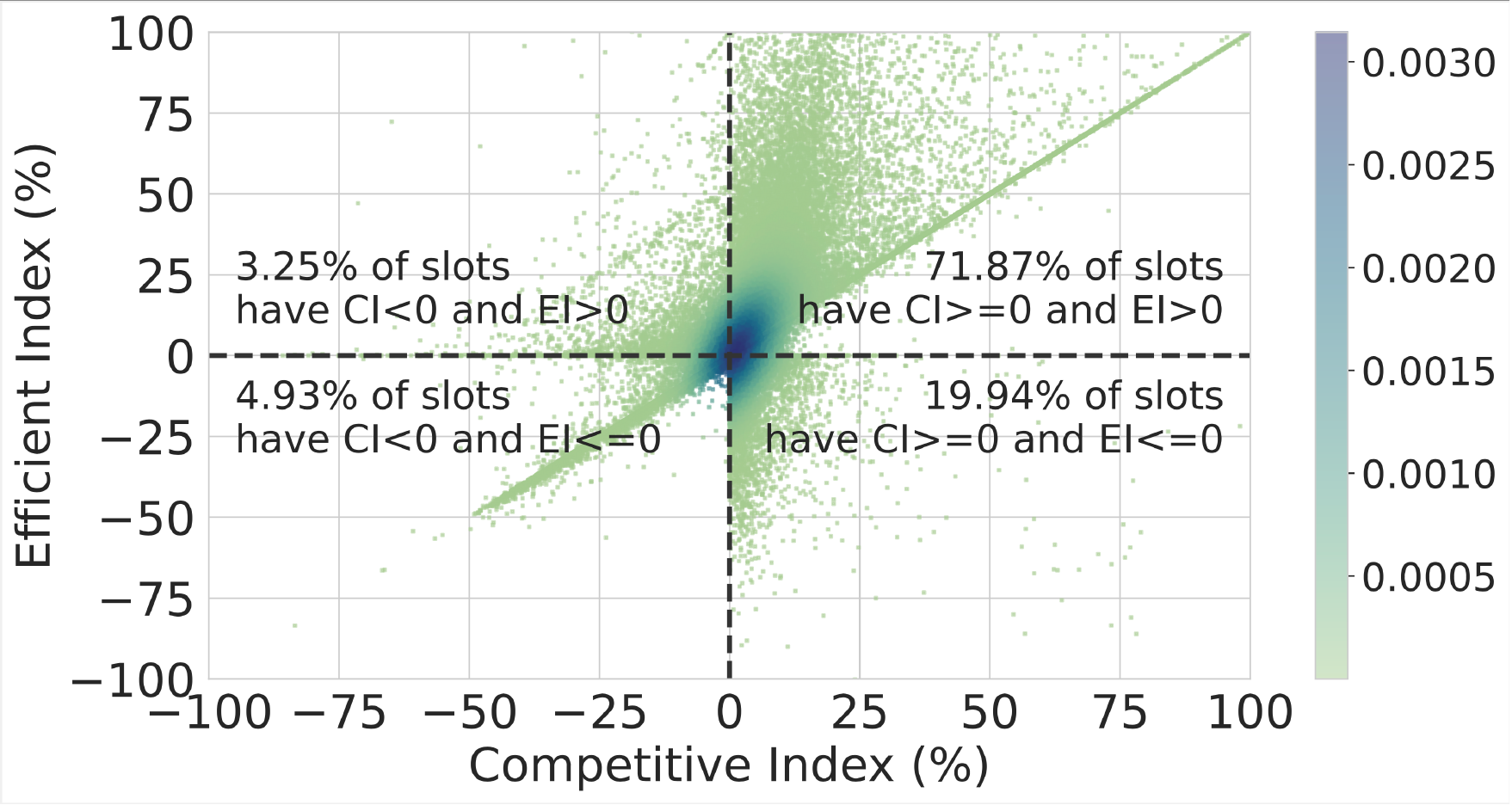}
    \caption{The distribution of CI and EI, where each dot represents an auction. The color represents density when multiple dots overlap.
    91.8\% of the MEV-Boost auctions are competitive (sum of the slots having $CI\geq0$), and 75.1\% of them are efficient (sum of the slots with $EI>0$). 
    Dots on the diagonal line ($CI=EI$) are auctions where the winners bid the true values. The dots above this line are auctions where the winners bid lower than true values.}
    \label{fig:epi}
\end{figure}

\parhead{Analysis of uncompetitiveness and inefficiency}

\autoref{fig:epi} shows the calculated CI and EI distribution across all MEV-Boost auctions in our dataset. Notably, 91.8\% (sum of the areas with $CI \geq 0$) of the auctions are competitive, yet only 75.1\% (sum of the areas with $EI > 0$) are efficient.

We note~\autoref{fig:epi} shows a somewhat strict definition of competitiveness (i.e., $CI \ge 0$). 
In reality, builders may bid slightly lower than their true value to ensure profits. 
Therefore, it is reasonable to relax and consider an auction with $CI \geq -\delta$ to be competitive for a small $\delta$. We call this notion $\delta$-competitive.
As shown in~\autoref{fig:ci-cdf}, when we consider $\delta= 1\%$ or $\delta=2\%$, the percentage of uncompetitive auctions reduces to about 6\% and 5\%, respectively. 
That is, under the slightly relaxed definition, approximately 95\% auctions were competitive.

\parhead{MEV's implication on auction outcomes}

\begin{figure}[tbp]
    \centering
    \includegraphics[width=\columnwidth]{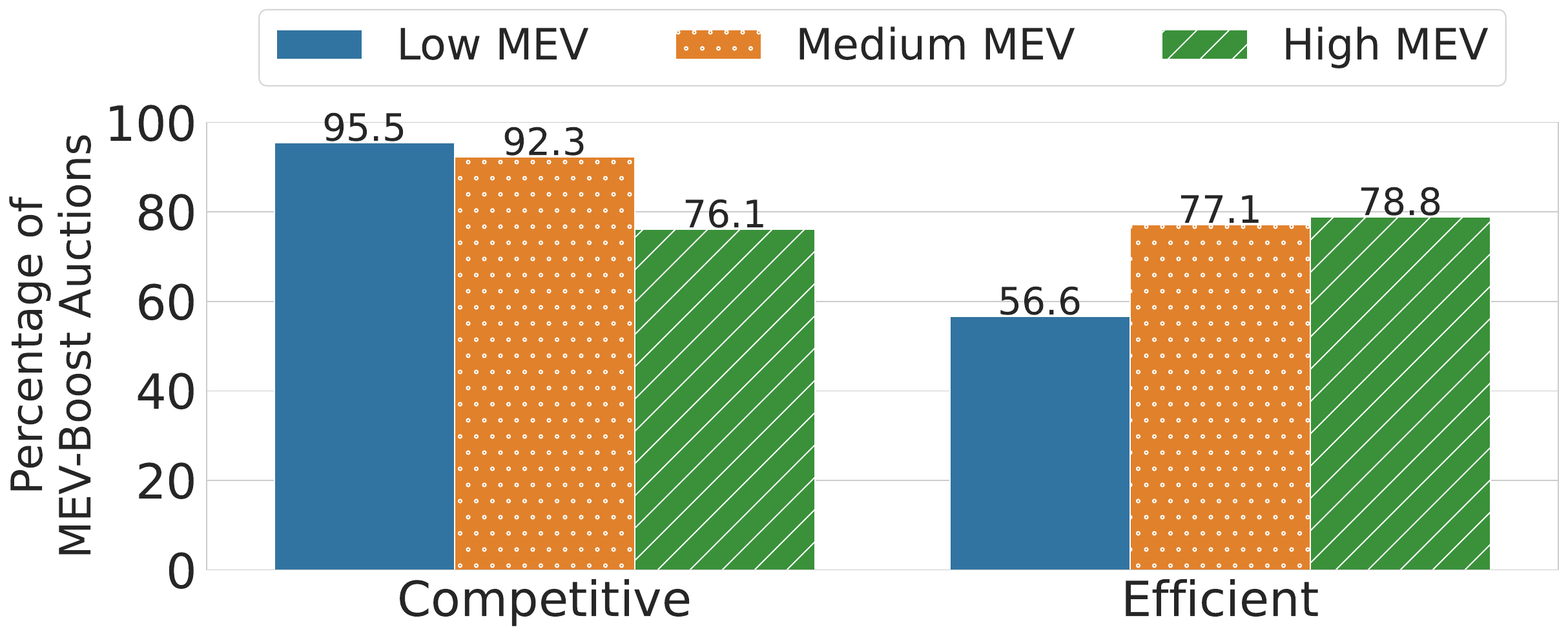}
    \caption{The ratio of auctions that are competitive and efficient across various MEV tiers.}
    \label{fig:epi-mev-range}
\end{figure}

To further understand how the magnitude of MEV in a slot affects the competitiveness and efficiency of the auction, 
we categorize every slot into one of three tiers based on the winning bid (as a proxy for MEV significance in that slot): {low MEV} (below 0.0215 ETH, the 10th percentile of the winning bid values), {high MEV} (above 0.3933 ETH, 95th percentile) and {medium MEV} in between. 
The ratio of competitive auctions in each tier is shown in~\autoref{fig:epi-mev-range}. We note that competitiveness decreases as the MEV of a slot increases.

\finding{
MEV-Boost auctions are generally competitive, while the competitiveness of MEV-Boost auctions deteriorates as a slot's MEV increases. 
}

In the interest of space, we leave the analysis of efficiency to~\appendixsubsecref{appx:efficiency}.

\parhead{Loss from uncompetiveness}
For an uncompetitive MEV-Boost auction in slot $s$, let $TV(p_2, s)$ denote the second-highest true value and $BV_w(s)$ denote the winning bid. The loss from uncompetitiveness is defined as
\[
L_{\text{uncomp.}}(s) = TV(p_2, s) - BV_w(s),
\]
representing the loss a proposer incurs compared to what it would have received in a competitive auction.
We computed this loss using our dataset, and the results are shown in~\autoref{tab:validator-loss-count}. The loss from uncompetitiveness ranges from 0.5\% to 1.7\% during our study periods.

\begin{table}[tbp]
    \centering
    \small
    \caption{Losses from uncompetitiveness and inequality.}
    \resizebox{\columnwidth}{!}{
    \begin{tabular}{l|p{1cm}|p{1cm}|p{1.6cm}|p{1.6cm}}
    \toprule
    \textbf{Time} & \textbf{Slots} &  \textbf{Profits (ETH)} & \textbf{$L_{\text{uncomp.}}$ (ETH) (\%)} & \textbf{$L_{\text{inequal.}}$ (ETH) (\%)} \\
    \midrule
      April 9-15  & 28,385 & 2,704.4 & 46.9 (1.7) & 312.1 (11.5) \\
      May 1-7   & 30,300 & 9,331.7 & 115.8 (1.2) & 518.6 (5.6) \\
      June 1-7  & 35,443 & 4,341.8 & 25.1 (0.6) & 342.2 (7.9)  \\
      July 1-7  & 36,040  & 3,938.8 & 19.1 (0.5) & 246.1 (6.3) \\
      August 1-7 & 17,831 & 2,135.5 & 12.5 (0.6) & 146.6 (6.9) \\
      September 1-7 & 35,815 & 3,939.5 & 48.5 (1.2)& 247.7 (6.3) \\
      October 1-7 & 39,131 & 3,427.0 & 28.9 (0.8) & 234.1 (6.8) \\
      November 1-7 & 39,848 & 4,736.2 & 45.6 (1.0) & 305.0 (6.4) \\
      December 1-7 & 38,686 & 5,884.4 & 80.0 (1.4) & 365.8 (6.2) \\
    \bottomrule
    \end{tabular}
    }
    \label{tab:validator-loss-count}
\end{table}

\subsection{Inequality in Block-building Capacities}
\label{sec:capability}
The previous section shows that the MEV-Boost auctions are competitive in most cases. However, a proposer may still incur loss due to a significant block-building capacity gap between builders.
In MEV-Boost auctions, this quantity is captured by the true values of the builder's bids.

\parhead{Quantifying the inequality of block-building capacities}

We compare the highest true values of different builders in the same slot to understand the differences in builders' ability to extract values. 
If a builder consistently has the highest true value, then she can potentially dominate the builder market. On the contrary, if many builders have similar true values, then it is plausible that a properly designed market could avoid a monopoly.

To quantify the disparity of block-building capacity, we compute the quartile coefficient of dispersion (QCD)~\cite{bonett2006confidence} of the highest true values in a given slot.
Specifically, suppose the true values of different builders in slot $s$ are \(\vec{tv}_s\). Let  \(Q_1(\vec{x})\) and \(Q_3(\vec{x})\) representing the first and third quartiles of \(\vec{x}\) (i.e., 25\% and 75\% percentiles). The QCD of slot $s$ is 
\[
QCD_s = QCD(\vec{tv}_s) = \frac{Q_3(\vec{tv}_s) - Q_1(\vec{tv}_s)}{Q_3(\vec{tv}_s) + Q_1(\vec{tv}_s)}.
\]

QCD is a real value in $[0,1]$. A larger QCD indicates greater dispersion (the intuition is that in a dataset with big dispersion, $Q_3$ is much larger than $Q_1$, thus $QCD$ approaches $1$). Since QCD uses quartiles, this metric is robust to outliers~\cite{bonett2006confidence}.

We now present the result given by QCD, noting that other metrics lead to a similar conclusion (see~\autoref{sec:appendix-measuring-other-metrics}).

\parhead{Evaluation results} 
We categorize the builders into three groups based on their market share from April to December 2023: top builders (top 5), middle builders (6-15), and tail builders (16-25). 
To understand how the magnitude of MEV in a slot affects the block-building capability, we follow the same methodology in~\autoref{sec:quantifying-auction} and categorize the auctions into three tiers: low MEV, medium MEV, and high MEV.

\begin{figure}[!tbp]
  \centering
    \includegraphics[width=\columnwidth]{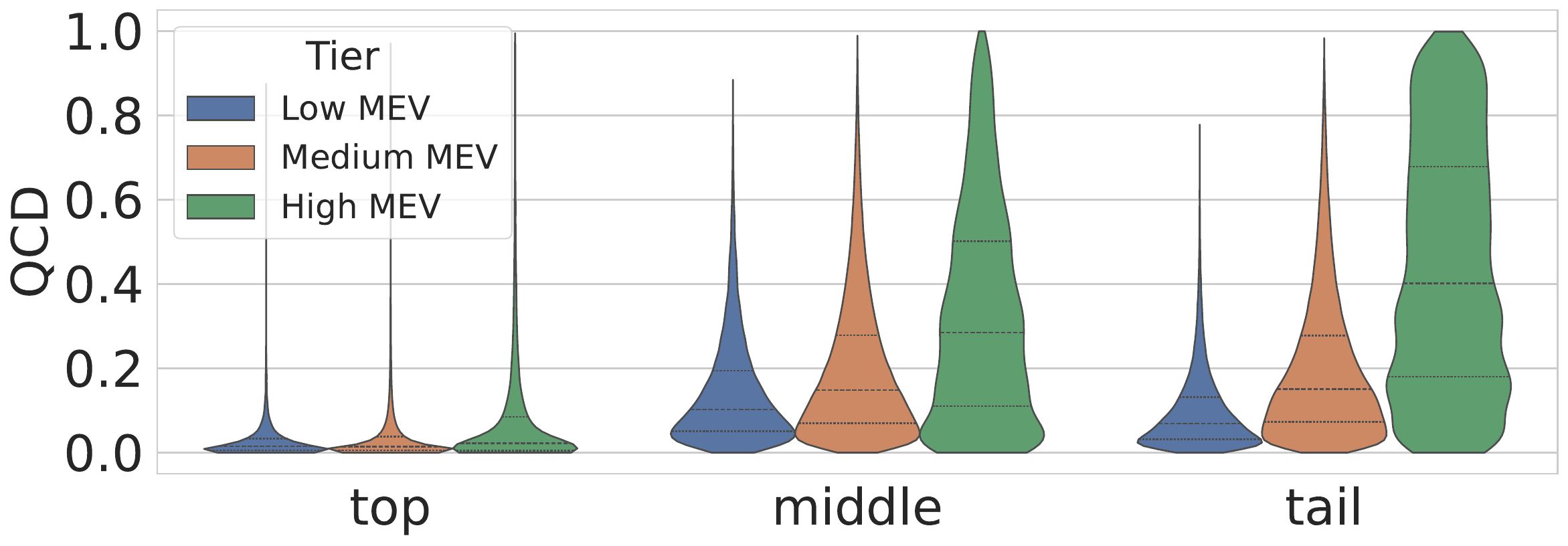}
  \caption{Violin graphs for the distribution of the QCD across three MEV tiers, for top, middle, and tail builders. The QCD measures the inequality of block-building capacity across builders. This figure reveals that such inequality worsens as MEV increases. Within a given MEV tier, inequality is greatest among tail builders, followed by middle builders, and is least among top builders.}
  \label{fig:true-value-qcd}
\end{figure}

In~\autoref{fig:true-value-qcd}, we plot the distribution of the QCD for MEV-Boost auctions from April 2023 to December 2023 for different builder groups and MEV tiers.
(We sample the first week of each month. See~\autoref{sec:data-collection} for details.)
It reveals two trends.
First, within each MEV tier, there is a clear increase in inequality (a high QCD means high inequality) from the top to the middle and then to the tail. This suggests that while top builders have comparable block-building capacities, middle and tail builders have significantly different capacities.
This might be good news: top builders having similar abilities means they can meaningfully compete, which is positive for the decentralization of the builder market.
Second, the inequality worsens as the MEV tier increases. 
One possible reason is that providers may give significant MEV opportunities to select builders, as we will discuss in the next section.
\done%

\finding{
The inequality in block building is lowest among top builders, higher among middle builders, and most pronounced among tail builders, and the inequality worsens as a slot's MEV increases.
}

\parhead{Loss from inequality}
For the auction in slot $s$, let $TV(p_1, s)$ and $TV(p_2, s)$ denote the highest and second-highest true values, respectively. The loss from inequality is defined as
\[
L_{\text{inequal.}}(s) = TV(p_1, s) - TV(p_2, s),
\]
reflecting the potential loss a proposer may face due to unequal block-building capacities, since a competitive auction only ensures the winning bid slightly exceeds the second-highest true value.
As shown in~\autoref{tab:validator-loss-count}, the loss from inequality ranges from 5.6\% to 11.5\% over the periods in our dataset.
Comparing the two types of loss, we find that inequality is the primary source of proposers' loss.

\subsection{Proposer Losses Over Extended Periods}

Note that the time range of the full bids is from April to December 2023, raising the natural question of whether proposers may incur more significant losses in 2024.

Recall that, in an MEV-Boost auction in the slot $s$, the proposer's loss is calculated as  $TV(p_1, s) - BV_w(s)$.
Unfortunately, without the full bid dataset, we cannot compute the highest true value $TV(p_1, s)$ and, therefore, the proposer loss.
However, since the winner's true value cannot exceed the highest true value (otherwise, it would be the highest itself),
we can compute a {\em lower bound} for the actual proposer losses using $TV_w(s)-BV_w(s)$, which we refer to as the {\em estimated proposer loss}. A significant estimated proposer loss suggests that proposers incur substantial losses.

\begin{figure}[t]
    \centering
    \includegraphics[width=\linewidth]{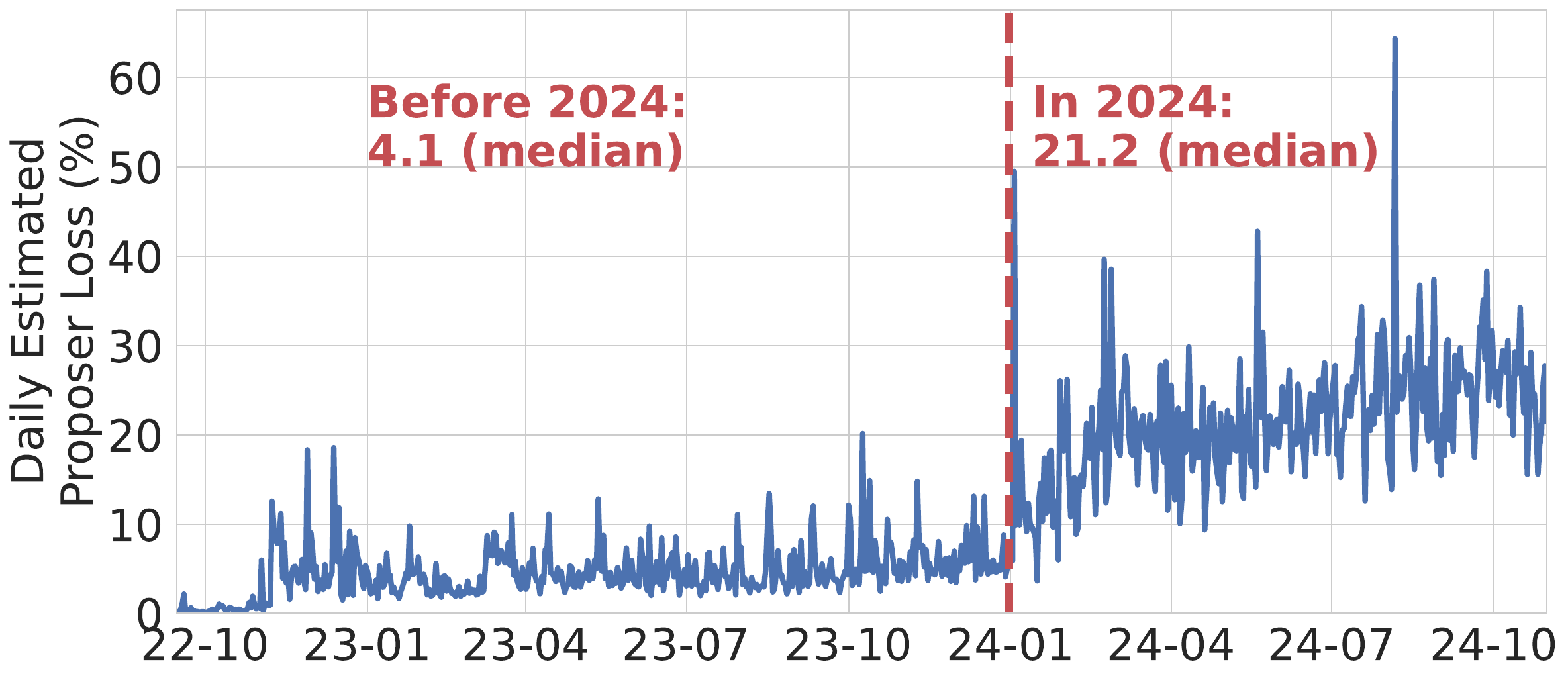}
    \caption{Daily estimated proposer loss over two years. Proposers suffered significant losses in 2024.}
    \label{fig:proposer-loss}
\end{figure}

We calculate the daily estimated proposer loss by aggregating the estimated losses from all MEV-Boost auctions each day and comparing this total to the overall block value for that day. As shown in~\autoref{fig:proposer-loss}, there is a clear upward trend in the daily estimated proposer loss in 2024---the median was 4.1\% before, compared to 21.2\% afterward! 
The daily estimated proposer loss is significant on August 6, 2024, primarily because, in an auction on that date (slot 9676257), the winner's bid is 11 ETH while the true value is 803 ETH---resulting in a significant loss for the proposer.

Why are proposer losses more significant in 2024?
One possible explanation is that worsening inequality in 2024 amplifies proposer losses. To explore this, we need to deepen our understanding of inequality in the builder market.

\section{Private Order Flows}
\label{sec:openness}

In the previous section, our study on historical auctions reveals that inequality is a primary source of proposers' losses. Recall that the inequality loss comes from the disparity in builders' true values---a bid's true value is computed by the total value extracted from the transactions.
As introduced in~\autoref{sec:mev-supply-chain}, transactions in Ethereum have two ways to reach a builder: through the public peer-to-peer network or via a direct channel. The former is publicly available to all, which is less concerning, while the latter is restricted, providing access only to select builders.

In this section, we further study the critical factor contributing to inequality---the difference in private order flows that builders receive.
\done%

\begin{figure}
    \centering
    \includegraphics[width=\columnwidth]{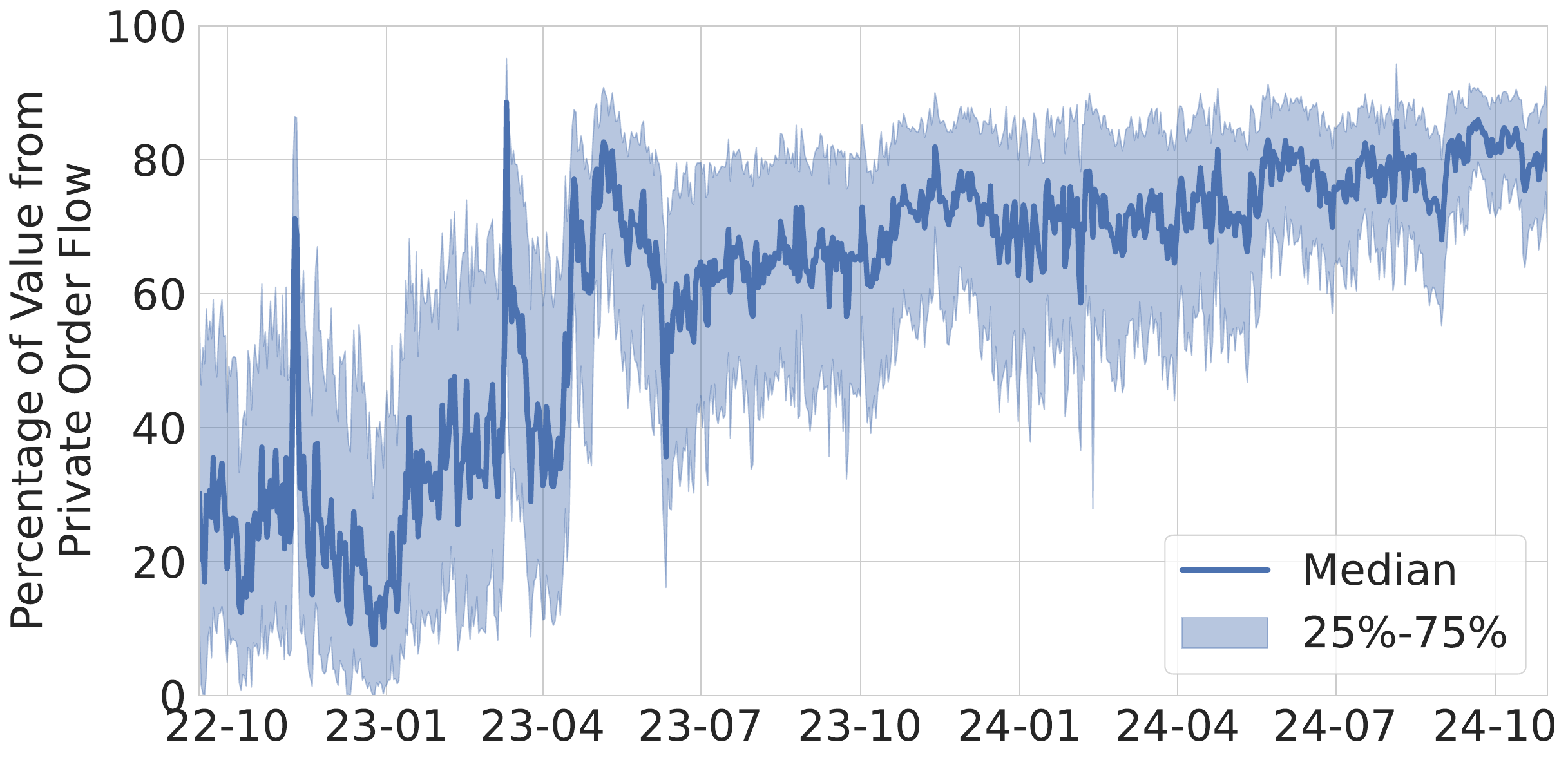}
    \caption{Fraction of block value from private order flows.}
    \label{fig:orderflow-percent}
\end{figure}

\parhead{Criticalness of private order flows}
To illustrate the importance of private order flow to builders, we compute the ratio of block values derived from private transactions.
We first identify all private order flows using data described in~\autoref{sec:data-collection}. Then, for each block, we compute the ratio between values extracted from private transactions (i.e., the total priority fees~\cite{EthereumGas} and direct transfers~\cite{flashbots_coinbase_transfer} to the block builder) to the block value.
\autoref{fig:orderflow-percent} plots the fraction of block value from private order flows over two years. As of October 2024, approximately 80\% of the block value comes from private order flows for the majority of the blocks, suggesting the criticalness of private order flows!

\finding{
Private order flows are critical to builders' block-building capacities since they contribute to more than 80\% of the block value in over 50\% of blocks. 
}

\subsection{Pivotal Private Order Flow Providers}
\label{sec:pivotal-providers}

Private order flows are critical to builders, and they can drive significant differences in builders' block-building capacities, contributing to inequality. Therefore, analyzing them is essential for understanding such inequality,
\done%
despite the challenges posed by the vast data volume and extended time span (112 million private transactions over two years).

Our first step is to classify the large volume of private order flows based on their sources. The entities who provide (or sell) private order flows to builders are called private order flow providers, or providers for short. Prominent private order flow providers are listed in~\autoref{tab:comparison-of-external-resources}.
Based on their roles in the MEV supply chain (see~\autoref{fig:background}), we categorize them into three types: \textit{searcher}, \textit{Telegram trading bot}, and \textit{channel}. The process of identifying the order flows from these providers, as well as the details of the finalized dataset, are described in~\autoref{sec:data-collection}.

Not every provider is equally important.
Some providers generate more value than others and their order flows can significantly impact builders. Therefore, our next step is to identify these important providers.

An existing approach~\cite{lu2023illuminating, OrderflowArt2024} is to compute the contribution of each provider to the builders' income (\autoref{sec:model}) during a specific period.
However, this approach can have false positives; for example, it could identify providers that were only influential temporarily.
\done%
For example, a single private transaction\footnote{\scriptsize  0xf0464b01d962f714eee9d4392b2494524d0e10ce3eb3723873afd1346b8b06e4} transferring 457.55 ETH to the builder as a tip could contribute significantly to the builder's income,
but this provider may only influence the MEV-Boost auction in a single slot.
Thus, we assess the significance of a provider as the number of auctions for which they played an important role; in particular, we focus on the providers that have sustained impacts (e.g., they can affect half of the MEV-Boost auctions in over two weeks).

We define a provider $P$ as {\em pivotal} for an instance of MEV-Boost auction if removing $P$'s transactions from the winning block causes the winner to lose the auction. Intuitively, had the winner not accessed $P$'s order flow, the winner might lose.\footnote{\scriptsize Note that the winner may or may not actually lose as it may be able to find an alternative pivotal provider, but this does not invalidate the fact that $P$ is pivotal.}
We identify pivotal providers in a given auction as follows: 
for each winning bid, we identify the providers of private order flow within the block and subsequently calculate the revised true value excluding each provider's contribution. If the recalculated true value drops below the next highest bid (from the partial bids dataset~\autoref{sec:data-collection}), the involved private order flow provider is considered pivotal.
Note that there can be multiple pivotal providers for an auction.

\begin{figure}[!tbp]
  \centering
    \includegraphics[width=0.97\columnwidth]{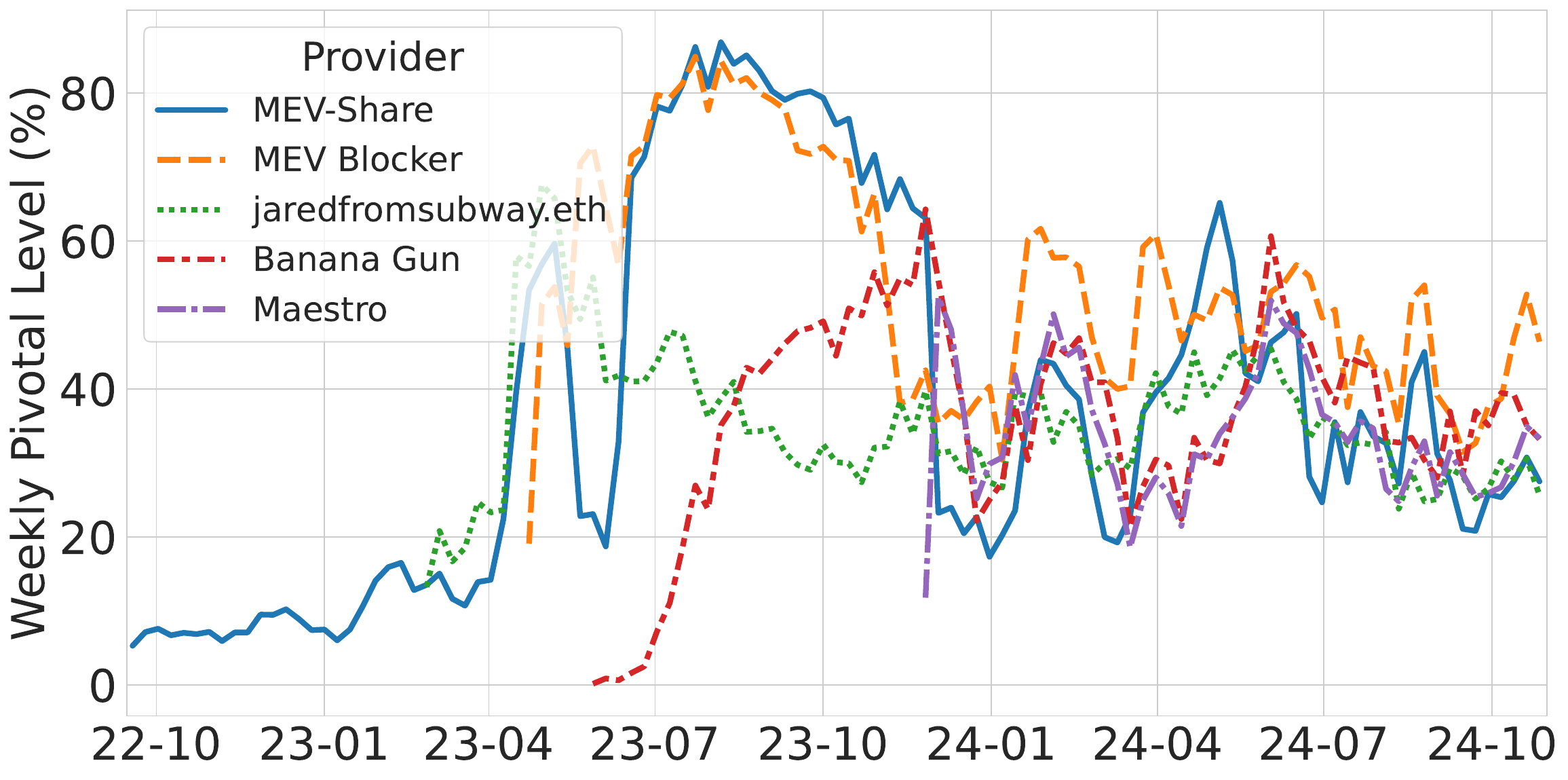}
  \caption{
  The pivotal level of top-5 pivotal providers over time. A provider $P$ is pivotal if the winner of an auction could have lost without private order flows from $P$. A line may not start from the beginning because the provider did not exist at that point. 
  }
  \label{fig:provider-of-crucial-transactions}
\end{figure}

\parhead{Pivotal level}
For each pivotal provider, we compute the fraction of auctions in which it is pivotal. We call this metric the {\em pivotal level}, a number from 0 to 1.
A higher pivotal level indicates a provider's greater significance in the MEV-Boost auctions.
\autoref{fig:provider-of-crucial-transactions} plots the pivotal level of the top-5 providers.
A builder who lacks access to any of their order flows would lose the majority of auctions.
However, order flows from the pivotal providers may not be accessible to all builders, which requires our further investigation.
We also make three observations from~\autoref{fig:provider-of-crucial-transactions}, which are presented in~\appendixsubsecref{appx:pivotal-providers} due to space constraints.

\parhead{Accessibility by multiple builders} \done%
Among the top five pivotal providers, MEV-Share and MEV Blocker are public-facing services with clear documentation and open communication channels. This enables us to analyze their statements regarding how their order flows are distributed to different builders---both claim that every connected builder receives the order flows~\cite{FlashbotsMEVShare,MEVBlocker}.
Although the order flows are not accessible to all builders (as summarized in~\autoref{tab:comparison-of-external-resources}, presented in \autoref{appx:subsidy}), top builders such as Beaver, Titan, and Rsync have access to them. Ideally, competition among these top builders will ensure that the value from these order flows is ultimately captured by the proposers.

Notably, while these order flows are distributed equally to all connected builders, inequality still exists.
Order flow providers may exclude less reputable builders due to trust concerns because builders are at an advantage and malicious builders can harm providers by, e.g., sandwich attacks~\cite{zhou2021high} and imitation attacks~\cite{qin2023blockchain}.
Although this exclusion may not harm proposers, it could exacerbate builder centralization, as new builders must expend significant effort to meet the requirements imposed by these pivotal providers.
Due to page constraints, we provide a more detailed discussion of the trust barrier and how it prevents builder decentralization in~\appendixsubsecref{appx:subsidy}.

\subsection{Exclusive Pivotal Providers}
Pivotal providers like MEV-Share and MEV Blocker send their order flows to connected builders who satisfy their requirements.
Many other providers (e.g., most searchers) do not publicly share their order flows; some only send order flows \textit{exclusively} to specific builders.
If multiple builders receive the same order flows, the proposer can, in principle, capture the value due to the competition. In contrast, a builder receiving exclusive order flows is detrimental to proposers, as it can exacerbate inequality and ultimately lead to losses for proposers.

The exclusive private order flows from a provider to a builder indicate a special relationship called {\em integration}~\cite{gupta2023centralizing}.
In this section, we further study the integration between builders and providers in the current market and how much loss proposers incur due to the integration.

\parhead{Case study of block 21,000,813}
To better understand exclusive order flows and their impact, we go through a case study of block 21,000,813, which has the highest estimated proposer loss in October 2024.
In the auction for this block, Titan received a valuable order flow from Banana Gun (55.5 ETH) and bid 2.8 ETH, while the second-highest bid was only 0.028 ETH. The significantly lower bids from other builders suggest they did not receive this order flow; otherwise, they could have bid closer to Titan.

In this case, the order from Banana Gun dominated the auction, as whoever had that order could win.
Thus, there is a clear {\em incentive} for the provider when its order flows are highly profitable: sending order flow to a select builder is more advantageous than distributing it to multiple builders.
\done%
The select builder only needs to bid slightly higher to win, allowing the provider and the builder to share the surplus MEV (52.7 ETH in this case).

\parhead{New integration pattern} The above case highlights a potential integration between Titan and Banana Gun. Previous studies~\cite{gupta2023centralizing,heimbach2024non} analyze an integration pattern between builders and searchers engaged in arbitrage between centralized exchanges (CEXs) and decentralized exchanges (DEXs).
These searchers may be run by the same party as their integrated builders and exclusively send order flows to them,  making the integrated builders more likely to win auctions when MEV from CEX-DEX arbitrage is significant.

Unlike known integrated searchers, Banana Gun is operated by a different party from Titan and typically sends order flows to multiple builders in most cases.
For example, in October 2024, Banana Gun's order flows were included in more blocks built by other builders than in those built by Titan.
This suggests that Titan may only receive exclusive order flows from Banana Gun when those flows are particularly profitable.
Similarly, other pivotal providers may tend to integrate with top builders when their order flows are highly profitable due to incentives.

To confirm this conjecture and evaluate the extent to which proposer loss is related to integration, we need to investigate exclusive order flows from providers to builders.

\begin{table*}[htbp]
\centering
\small
\resizebox{\textwidth}{!}{
\begin{threeparttable}
\caption{The top 10 providers sending exclusive order flows and their corresponding builders, contribute to the highest estimated proposer loss (before and in 2024).}
\label{tab:comparison-of-loss}
    \begin{tabular}{c|c|c|r|r|c|c|c|r|r}
        \toprule
    \multicolumn{5}{c|}{\textbf{before 2024}} & 
    \multicolumn{5}{c}{\textbf{2024}} \\
    \cmidrule{1-5} \cmidrule{6-10}
    \textbf{Provider} & 
   \textbf{MEV Activity} & 
    \textbf{Builder} & \textbf{Blocks (\#)}  & \textbf{Losses (ETH) (\%)}
     &  \textbf{Provider} & 
   \textbf{MEV Activity} & 
    \textbf{Builder} & \textbf{Blocks (\#)}  & \textbf{Losses (ETH) (\%)}\\
        \midrule
SCP & CEX-DEX arbitrage & Beaver & 42,639 & 4,566.0 \ (22.7) & Banana Gun & Sniper & Titan  & 34,442 & 17,488.1  \ (32.6)  \\
Searcher: 0x57c1... & CEX-DEX arbitrage & Addr: 0x4737... & 12,220 & 1,088.7 \ (5.4) & SCP & CEX-DEX arbitrage & Beaver & 13,148 & 3,077.8 \ (5.7) \\
Wintermute & CEX-DEX arbitrage & Rsync & 13,823 & 642.6 \ (3.2) & Searcher: 0x6980... & Atomic arbitrage & I can haz block? & 516 & 2,495.7 \ (4.7) \\
Searcher: 0xf8b7... & CEX-DEX arbitrage  & Manta & 9,699 & 473.9 \ (2.4) & jaredfromsubway.eth & Sandwich & Beaver & 22,230& 2,480.5 \ (4.6) \\
Searcher: 0x9dd8... & Atomic arbitrage & Addr: 0x3bee... & 1,826& 424.9 \ (2.1) & Searcher: 0x6454... & Liquidation & Addr: 0x3bee... & 809 & 1,231.8 \ (2.3) \\
Searcher: 0x761d... & Atomic arbitrage & builder0x69 & 19 & 287.5 \ (1.4) & WINTERMUTE & CEX-DEX arbitrage & Rsync & 8,586 & 1,038.3 \ (1.9)  \\
Searcher: 0x6454... & Liquidation & Addr: 0x3bee... & 456 & 246.6 \ (1.2) & c0ffeebabe.eth & Liquidation & Ty For The Block & 368& 1,003.6 \ (1.9) \\
Searcher: 0x5e51... & Atomic arbitrage  & eth-builder & 370 & 226.5 \ (1.1) & Banana Gun & Sniper  & Beaver  &  2,424 & 920.5 \ (1.7) \\
Searcher: 0xeeaa... & Atomic arbitrage & eth-builder & 67 & 195.6 \ (1.0) & Maestro &  Sniper & Beaver & 10,587 & 439.7 \ (0.8) \\
Searcher: 0xe20c... & Atomic arbitrage & Addr: 0x5638... & 592 & 168.9 \ (0.8) & Searcher: 0x77d5... & Atomic arbitrage & Rsync & 797& 333.3 \ (0.6) \\
        \bottomrule
    \end{tabular}
\end{threeparttable}
}
\end{table*}

\parhead{Challenges of identifying exclusive order flows}
Identifying exclusive order flows is challenging, as we do not know the exact order flows each builder receives.
The on-chain blocks only record the winning bids of MEV-Boost auctions, which reflect the order flows received by the winners.
Even though we have the full bids dataset from 2023, it remains unclear whether the builders did not receive an order flow or chose to exclude it.
Analyzing the number and value of transactions sent from providers to builders is also unreliable because providers might selectively send exclusive order flows, and the difference may not be very significant.

Existing work~\cite{gupta2023centralizing} do not provide a methodology for identifying exclusive order flows, while~\cite{heimbach2024non} identifies integration by analyzing the correlation between the volume of searchers' CEX-DEX arbitrage and the builders' daily market share of blocks.
This approach cannot be directly applied to identify generalized integration between providers and builders, as different providers may perform various MEV activities beyond CEX-DEX arbitrage.

Most importantly, since we are interested in how many proposer losses are related to integration, identifying a builder-provider integration pair alone cannot accurately answer this question.
\done%
As discussed earlier, a provider may send order flows exclusively to one builder under specific conditions. \done%
Therefore, attributing all proposer losses in a builder's blocks to its integration might be inaccurate.

\parhead{Our approach}
Despite the challenges mentioned above, we are still able to infer whether an order flow is likely to be exclusively sent to a builder based on a reasonable assumption: all builders are rational and do not collude, so they will compete upon receiving the same order flow. We justify this assumption with our study in~\autoref{sec:competition}, which shows that over 90\% of MEV-Boost auctions are competitive.
Based on this assumption, we design an algorithm to infer if an order flow is exclusively sent to a builder.

For a given slot, we obtain the winning bid and compute the second-highest bid from our partial auction dataset. Then, for the winning bid in this slot, we identify the private order flows within this block and their associated providers using our private transaction dataset.
For each provider, we evaluate whether the combined value of the private order flow from this provider and the public order flow exceeds the second-highest builder's bid multiplied by a threshold. This threshold, set at 1.15, was derived from our analysis of historical MEV-Boost auctions, where less than 5\% of builders withheld bids that exceeded this value. The threshold accounts for the possibility that builders may bid slightly below their true valuations.
\done%
If this condition holds, it suggests that the provider exclusively sends order flows to the builder, thus giving the winner a competitive advantage.

Since there is no ground truth for exclusive order flows, we validate our method using four builder-searcher integrations identified in previous work~\cite{heimbach2024non}. %
Our algorithm cannot identify exclusive order flows when their value is insignificant, i.e., when they cannot determine the auction outcome. Thus, our approach will have false negatives.
\done%
We focus on precision, specifically assessing, for each provider, the percentage of exclusive order flows correctly attributed to their well-known integrated builder.
\done%

For the known builder-searcher integrations---Beaver with SCP, Rsync with Wintermute, Manta with searcher 0xf8b7..., and anonymous builder 0x4737... with searcher 0x57c1...---our methodology achieves accuracies of 99\%, 96\%, 99\%, and 99\%, respectively.
Except for Rsync and its integrated searcher, our methodology achieves very high precision: less than 1\% of order flows are inaccurately identified as exclusive (false positive). Further investigation into the identified exclusive order flows of Wintermute shows that about 70\% of the false positives occurred in blocks built before Rsync began block-building.
At this point, Rsync might send order flows to multiple builders. However, only one builder included the order flow before the auction concluded. Besides the known integrations, we also identified previously unknown integrations involving the top-5 pivotal providers---Banana Gun, Maestro, and jaredfromsubway.eth---as listed in~\autoref{tab:comparison-of-loss}.

\done%

\parhead{Proposer loss due to integration}
For each builder-provider integration pair identified by our algorithm, we count the number of blocks where the builder exclusively receives the order flows and calculate the total estimated proposer loss in these blocks.
\autoref{tab:comparison-of-loss} lists the top 10 providers and their corresponding builders receiving exclusive order flows, ranked by the resulting losses before and in 2024 respectively.
We classify each provider's MEV activity using previous studies~\cite{gupta2023centralizing,heimbach2024non} and MEV analysis tools~\cite{searcherbuilder2024, LibMEVLeaderboard2024}. For simplicity, we label each searcher with only its predominant MEV activity.
Telegram bots, typical sniper bots~\cite{cernera2023token}, automatically assist users in quickly buying new tokens at launch by monitoring liquidity pools.

We find that the top four exclusive providers causing the highest proposer loss before 2024 are integrated searchers engaged in CEX-DEX arbitrage.  This observation aligns with previous studies on competitive advantages and the profits associated with integration between these searchers and their integrated builders~\cite{gencer2018decentralization,heimbach2024non}, highlighting our algorithm's effectiveness in identifying exclusive order flows. \done%

Compared to the exclusive order flows before 2024, a more concerning problem is the formation of more exclusive relationships between top builders and top providers. This also confirms our conjecture. For instance, Banana Gun exclusively sends order flows to Titan in approximately 34K blocks, accounting for 32.6\% of the estimated proposer losses covered in our study. Additionally, two other top-5 pivotal providers, jaredfromsubway.eth and Maestro, also exclusively send order flows to Beaver, contributing to 5.4\% of the losses.
In contrast, the total losses caused by the integrated CEX-DEX arbitrage searchers remain relatively similar before and during 2024. Notably, searcher 0x57c1... and searcher 0xf8b7... stopped operating in 2023.
These results effectively explain that compared to the previous known integration, the emergence of new integration accounts for the majority of the proposer losses in 2024.

\finding{Top pivotal providers Banana Gun, jaredfromsubway.eth, and Maestro integrate with the top two builders, Titan and Beaver. Their integration causes significant losses to proposers.}

\parhead{Competing providers}
In addition to noting that three of the top five pivotal providers have started integrating with Beaver or Titan, \done%
we also observe that providers engaging in the same MEV activities tend to avoid integrating with the same builder simultaneously (Banana Gun sent order flows exclusively on Beaver in January 2024, while its flows to Titan occurred after this period).
A possible explanation is concern over trust: if a provider knows that a builder collaborates with its competitor, it may worry about the security of its order flow sent to that builder. The builder could potentially leak information about the provider's order flow to the competitor.

\section{Discussion}
\label{sec:discussion}

\subsection{Implications of Integration}
Beyond proposer loss, the integration carries additional implications, which we discuss below.

\parhead{Inaccurate MEV analysis} Our study reveals a potential pattern between providers and builders: when the order flows from a provider may determine the auction outcome, the provider exclusively shares the order flow with a specific builder.
In return, the provider may receive a refund from the builder. However, this refund could be off-chain, making it difficult to measure the providers' actual costs.  This limitation could lead to inaccuracies in MEV analysis tools. For example, the profit of a top MEV searcher, \textit{jaredfromsubway.eth}, is only 1.3\% of its total revenue according to~\cite{LibMEVLeaderboard2024}; its actual profit may be higher due to potential refunds.
  
\parhead{Suboptimal block-building} 
In principle, a builder with access to more profitable order flows can achieve better block-building.
However, integration between top providers and builders can prevent builders from maximizing potential MEV, resulting in suboptimal block-building in the winners' blocks and ultimately reducing profits for providers, builders, and proposers.
For example, Beaver's integrated searcher may identify a good opportunity for CEX-DEX arbitrage, while Banana Gun may also handle a large volume of private order flow exclusively sent to Titan.
These two order flows are not in conflict; however, due to the builders' competition, only one can be included in the winning block within the same slot. The excluded flow may not be included in the next block because the MEV opportunity could be time-sensitive (e.g., CEX-DEX arbitrages).

\parhead{Censorship}
A security concern of builder centralization is transaction censorship. Ethereum's current censorship resistance relies heavily on the choices of builders, particularly top builders. For example, before Beaver stopped censoring in July 2024, over 50\% of blocks were built by censoring builders~\cite{censorshippics}.
Our findings suggest that builder centralization remains difficult to overcome due to integration, potentially exacerbating the censorship problem on Ethereum.

\subsection{Proposed Mitigations and Their Limitations}
\label{sec:future-works}

\IDEA{
We now discuss why the proposed mitigation fails to resolve the identified problems. This highlights that addressing the identified problems is a grand challenge that requires further, possibly interdisciplinary, study.

\parhead{Removing trust barriers}
There is a lack of a trustworthy mechanism to facilitate fair exchanges between private order flow providers and builders.
As a result, some providers require builders to have a minimal market share, and we also observe that competing providers typically do not integrate with the same builder.

One promising direction to reduce this trust dependency is the use of encrypted mempools~\cite{charbonneau_encrypted_mempools}, where transactions remain encrypted until they are irreversibly included in a block, preventing builders from having complete knowledge of user transactions. For example, this can be implemented using Trusted Execution Environments (TEEs)~\cite{costan2016intel} (e.g., SUAVE~\cite{flashbots2024future}), though standard challenges such as side channels and covert channels~\cite{flashbots2023backrunning} still apply. 
\done%

Besides, requiring builders to achieve fair ordering~\cite{kelkar2020order,kelkar2023themis}, is another viable solution to prevent transaction manipulation, though a first-come-first-served ordering policy may introduce other problems like off-chain latency wars~\cite{mamageishvili2023buying}.
\done%

\parhead{Proposed PBS designs}
The Ethereum community has considered multiple alternative designs to mitigate the centralization and security problems in the out-of-protocol PBS. Enshrined-PBS (ePBS)~\cite{damato2024eip7732} is proposed to decentralize PBS auctions by removing centralized relays from PBS.
MEV burn, an add-on of ePBS, which helps smooth and redistribute MEV spikes, benefits all ETH holders by burning the winning bid value in the PBS auction~\cite{mevburn2024}. 
More recent proposals, including Execution Ticket~\cite{executiontickets2023} and Execution Auction~\cite{monnot2024proposers}, aim to further mitigate the effects of MEV on validator centralization by removing execution block proposing rights from validators.

However, in both eBPS and MEV burn, blocks are still produced through auctions, and the PBS auction can continue to exist as a secondary market in both Execution Ticket and Execution Auction. This allows the re-sale of block proposing rights to builders who can build more profitable blocks~\cite{executiontickets2023}.
None of these proposals considers the incentives of providers, so the integration problem in PBS auctions remains unresolved.

\parhead{Future work}
Based on the discussion, we identify that an effective mitigation must satisfy at least the following two properties:
First, it must provide a security guarantee to providers' order flows from potential attacks.
Second, it should consider providers' incentives, such as offering a game-theoretical guarantee to ensure they no longer have the motivation to continue such integration.
Identifying these two properties is a step forward, but fully addressing them requires significant further study for future work.
}

\subsection{Limitations of Our Approach and Data}
We compute the true values of bids following a standard approach as~\cite{eigenphi2023, flashbots_mev_inspect_py,heimbach2023ethereum, soispoke2023empirical}. This approach relies on identifying builder-controlled addresses, and missing addresses could cause inaccuracies. We cross-validate with multiple sources to ensure collected addresses are the most comprehensive to date (see~\autoref{sec:data-validation}).
It may underestimate values by ignoring off-chain payments, or overestimate profits if builders refund providers off-chain.
A builder with large off-chain payments could bid higher than the computed true value, appearing as an extremely large subsidy.
This could lead to the misclassification of an efficient auction as inefficient due to subsidization.
However, we observe that there are no excessively large subsidies, and most builders bid close to our computed true values in our study of competitiveness and inequality, suggesting that off-chain payments are not significant. 
While underestimating the winner's true value does not affect our competitiveness assessment, underestimating the second-highest true value could cause an uncompetitive auction to be misclassified as competitive. Nevertheless, this scenario is rare. In practice, builders with higher true values would have an incentive to bid higher to win, making such misclassifications unlikely.

The potential refunds to providers do not affect the computation of proposers' losses because this portion is not allocated to proposers, regardless of the volume of refunds. However, overestimating the second-highest true value could lead us to misclassify a competitive auction as uncompetitive. Our study shows that large-scale integration between top builders and providers---where such refunds are more likely---began in 2024, after the period covered by our evaluation of competitiveness and efficiency.

We might also miss pivotal providers due to incomplete coverage of transaction sources. For example, BloXroute (a widely used channel) does not publish the list of transactions they processed. As a result, some providers may be missing from our analysis. However, there will be no false positives because each provider attribution is either directly from official sources or through known searcher addresses (see~\autoref{sec:dataset}). This limitation leads to an underestimation of the significance of providers, but not an overestimation. Similarly, because partial bids are self-reported by relays, underestimating the second-highest bid could cause us to underestimate a provider's pivotal level.

\section{Related Work}
\label{sec:related-work}

\parhead{Empirical study on Ethereum's builder market}

Several previous works studied the centralization of the builder market shares.
An early exploration by Yang et al.~\cite{yang2022sok} analyzed the market share of relays and builders.
Their findings indicated a centralization of Ethereum's builder market, with Flashbots and Builder0x69 accounting for over 53\% of the market share between September 15th, 2022, and November 30th, 2022.
Similar studies on the builder market conducted by~\cite{heimbach2023ethereum,wahrstatter2023time} also confirmed the centralization in the builder market. Their observations were based on the market share and the distribution of total profits generated within the PBS ecosystem.
Moreover, several online dashboards~\cite{mevboost2024,relayscan2024} offer a dynamic view of the ecosystem's trends by presenting real-time PBS analytics. 
However, these studies do not quantify the causes of centralization or the security implications of a centralized builder market.

\parhead{Private order flow}
Many studies also discuss the importance of private order flows, particularly focusing on private order flows from integrated searchers.
Thiery~\cite{soispoke2023empirical} counted the total number of transactions, along with their values, from June 1 to July 15, 2023. He found that private order flows account for only 30\% of the transactions yet represent 80\% of the total value paid to builders. 
Our study highlights a similar result: private order flows account for more than 80\% of the block value in most blocks.
Gupta et al.~\cite{gupta2023centralizing} confirmed that builders are more likely to win the MEV-Boost auctions when their integrated searchers can provide high-value exclusive order flow by effectively exploiting CEX-DEX arbitrage.
Heimbach et al.~\cite{heimbach2024non} observed that during their study, Beaver received a total of 1,941.1 ETH in transaction fees but paid proposers 6,620.94 ETH. This large gap suggests that Beaver might have received significant off-chain profits from its integrated searchers.
Öz et al.~\cite{oz2024wins} employed Linear Discriminant Analysis to identify the integration between builders and providers.
Our study uncovers previously unknown integration of top providers and quantifies the resulting proposer losses.

\parhead{Strategic behaviors in MEV-Boost auctions}
The strategies used by different entities in the MEV-Boost auctions and their potential impacts have attracted significant interest from researchers.
The strategic behaviors of proposers were first studied by~\cite{wahrstatter2023time,schwarz2023time} focusing on the timing game where proposers strategically delay block proposals to maximize their profits.
Öz et al.~\cite{oz2023time} further conducted an agent-based simulation to analyze how waiting games affect consensus stability, finding that a delay strategy can be profitable and not degrade consensus if sufficiently many validators adopt it.
Wu et al.~\cite{wu2024strategic} proposed a game-theoretic model for MEV-Boost auctions, using simulations to examine various builders' bidding strategies, including naive, adaptive, last-minute, and bluff bidding. Their results highlighted the importance of latency and exclusive order flow in the effectiveness of bidding strategies. 
Pai et al.~\cite{pai2023structural} analyzed the latency advantage of searcher-builder integration in adjusting bidding values.
The strategies discussed in these papers highlight the potential factors that can make MEV-Boost auctions uncompetitive or inefficient, while our paper quantifies these uncompetitive and inefficient auctions.

\parhead{Causes and implications of centralized builder market}
Previous works such as~\cite{gupta2023centralizing,kilbourn2022orderflow1} theoretically analyzed the potential centralizing effects of exclusive order flows on the builder market: builders with exclusive order flows are more likely to win MEV-Boost auctions, thus gaining further advantages in receiving order flows from non-integrated searchers.
Capponi et al. identified this dynamic as leading to a ``Proof-Of-MEV'' paradigm~\cite{capponi2024proposer}, where the capacity to capture MEV becomes a critical factor for success in MEV-Boost auctions.
The decentralization of Ethereum's builder market not only benefits itself but also positively impacts the decentralization of validators. A recent study showed that heterogeneity among validators in building blocks may lead to centralization; however, a decentralized and competitive builder market can encourage validator decentralization~\cite{bahrani2024centralization}. We complement these theoretical analyses with empirical studies to analyze the causes and potential problems of the current centralized builder market. %

\FULL{
\section{Conclusions}
In this paper, we challenge the community's belief that builder centralization is acceptable by empirically quantifying the significant profit losses incurred by validators in the current builder market.
These significant losses could lead to PBS's instability and MEV oracles' inaccuracy and, if left uncontrolled, may lead to validator centralization.
We attribute the loss to two factors: insufficient competition and inequality of block-building capacities, with the latter being the dominant reason.
We further investigated the incentive issues of the current MEV supply centralization that contribute to the builder centralization and significant proposer loss.
Finally, we identify the properties required to address these issues and hope these insights will benefit future work.

\section*{Acknowledgements}
This work is partially supported by an Ethereum Foundation Grant. 
We thank our shepherd and the reviewers for their helpful comments, which significantly improved the paper.
We are grateful to \ultrasound for sharing the full bid dataset.
We also thank the teams at BloXroute, Merkle, MeowRPC, MEV Blocker, Flashbots, and Blocknative for answering our questions regarding access requirements. 
Finally, we appreciate the insightful comments from Prof. Campbell Harvey and the valuable discussions with Thomas Thiery, Barnab\'{e} Monnot, Danning Sui, Burak {\"O}z, Data Always, and Georgios Konstantopoulos.
}

\bibliographystyle{IEEEtran}
\bibliography{ref}

\appendices
\section{Visualization for Distribution}
\label{sec:appendix-figures}

\begin{figure}[thbp]
    \centering
    \includegraphics[width=\columnwidth]{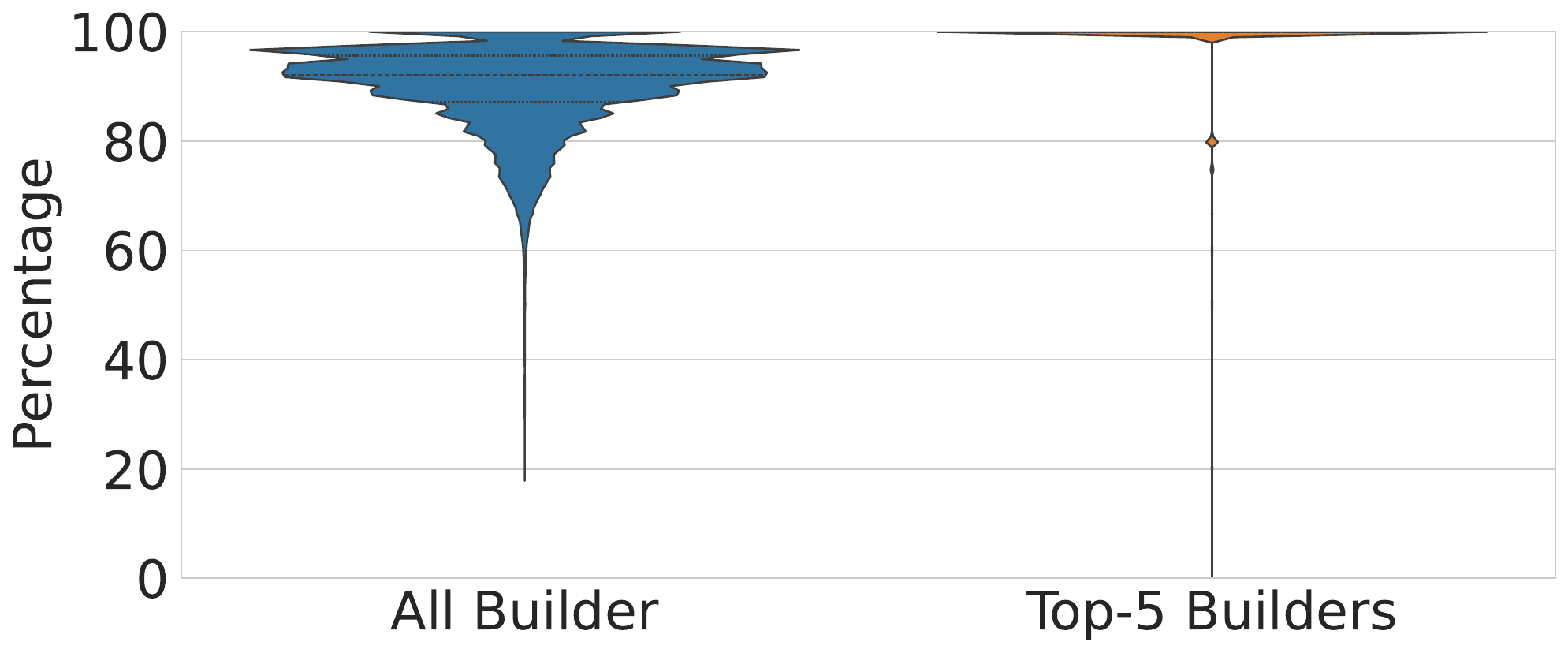}
    \caption{Distribution of the percentage of builders who submit to the \ultrasound.}
    \label{fig:bids-validation}
\end{figure}

\begin{figure}[thbp]
    \centering
    \includegraphics[width=\columnwidth]{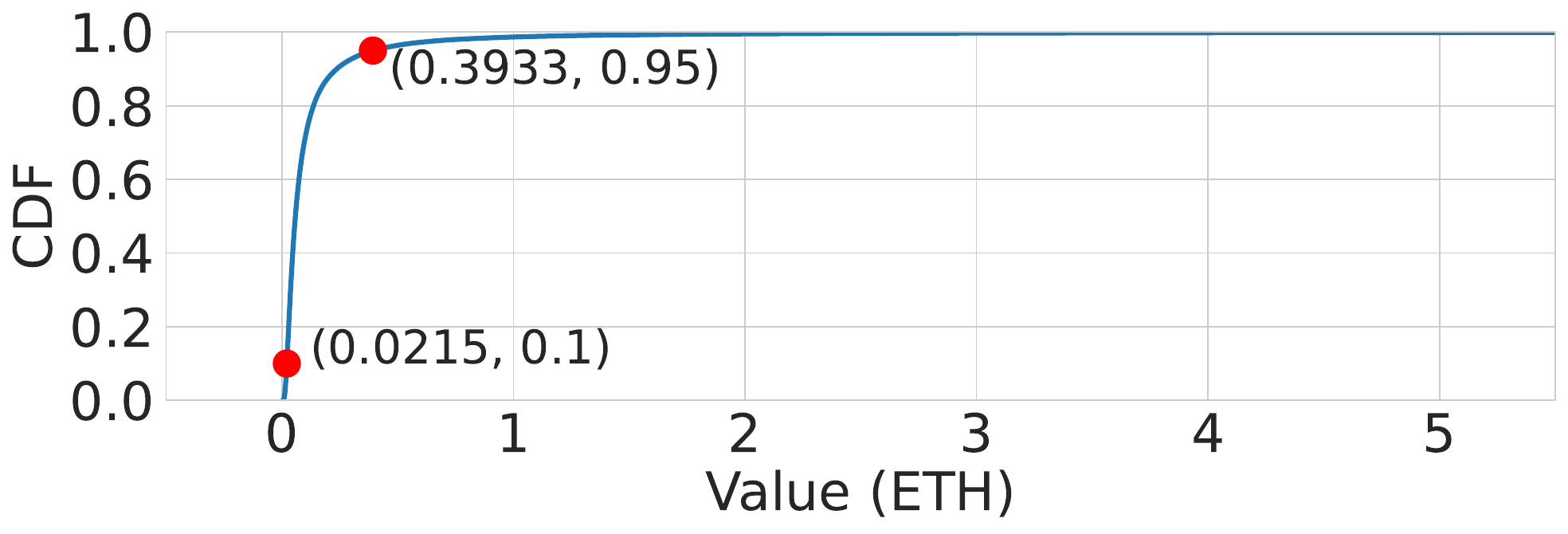}
    \caption{CDF of the winning bid value of auction dataset.}
    \label{fig:cdf-of-winning-bid-value}
\end{figure}

\autoref{fig:bids-validation} shows the distribution of the percentage of builders who submit to the \ultrasound.
\autoref{fig:cdf-of-winning-bid-value} shows the CDF of the winning bid values of MEV-Boost auctions.

\section{Data Collection}
\label{appx:data-collection}

\parhead{Builder public keys}
We collect builder public keys from three data sources. First, builders often disclose their public keys in official documents.
For example, Titan builder has disclosed 15 public keys in its official document~\cite{titanbuilder2024pubkeys}.
Second, some builders leave their unique marks in the blocks' ``extra data'' field. For example, Flashbots builder uses the mark \texttt{``Illuminate Democratize Distribute''}.
We manually group builder public keys based on these marks and exclude the public keys which are known as impersonation~\cite{titanbuilder2023blogpost}.

\parhead{Transaction sources} 
We identified transactions originated from top Telegram trading bots, including Maestro~\cite{Maestro2023}, Banana Gun~\cite{BananaGun2023}, and Unibot~\cite{UnibotX2023}, by searching for transactions that interacted with their official smart contracts.

Two channels, MEV Blocker and MEV-Share publish data~\cite{cowprotocol2024mevblocker,flashbots2024data} about transactions submitted to them.
Specifically, transactions submitted to MEV Blocker are publicly available on Dune~\cite{cowprotocol2024mevblocker}, from which we extracted 28,233,223 transaction hashes.
We obtained 24,449,086 transactions submitted to MEV-Share from the public Flashbots dataset~\cite{flashbots2024data}. 
These transaction labels are sourced from their official datasets and we view them as ground truth.

\section{List of Searchers}

In~\autoref{tab:searcher-list}, we provide the list of searchers and their addresses mentioned in our study.

\begin{table}[htbp]
\centering
\small
\caption{The list of searchers}
\resizebox{\columnwidth}{!}{
\begin{tabular}{ll}
\toprule
\textbf{Searcher} & \textbf{Address}  \\
\midrule
jaredfromsubway.eth & 0xae2Fc483527B8EF99EB5D9B44875F005ba1FaE13 \\
\midrule
\multirow{7}{*}{SCP} & 0xA69babEF1cA67A37Ffaf7a485DfFF3382056e78C \\
& 0x56178a0d5F301bAf6CF3e1Cd53d9863437345Bf9 \\
& 0xa57Bd00134B2850B2a1c55860c9e9ea100fDd6CF \\
& 0x4Cb18386e5d1F34dC6EEA834bf3534A970a3f8e7 \\
& 0x4Cb18386e5d1F34dC6EEA834bf3534A970a3f8e7 \\
& 0x5050e08626c499411B5D0E0b5AF0E83d3fD82EDF \\
& 0xFA103c21ea2DF71DFb92B0652F8B1D795e51cdEf \\
& 0x0DA9d9eCeA7235c999764e34F08499cA424c0177 \\
\midrule
\multirow{3}{*}{WINTERMUTE}
 & 0x0087BB802D9C0e343F00510000729031ce00bf27 \\
~ & 0x280027dd00eE0050d3F9d168EFD6B40090009246 \\
~ & 0x51C72848c68a965f66FA7a88855F9f7784502a7F \\
\bottomrule
\end{tabular}
}
\label{tab:searcher-list}
\end{table}

\section{Efficiency and Block-building Capacities}

\subsection{Efficiency of MEV-Boost Auctions}
\label{appx:efficiency}

\parhead{Analysis of inefficiency}\autoref{fig:epi} shows that 24.9\% of MEV-Boost auctions were not efficient. 
An auction is inefficient when the builder with the highest true value does not win. There are two possible reasons. 
First, a builder can win the auction without possessing the highest true value by overbidding, known as subsidization~\cite{relayscan2024}.
Second, builders may ``shade'' their bids (bid shading means strategically placing bids slightly below one's true value); if the builder with the highest true value ``shades too much,'' it will lose.

We analyzed the underlying reasons for all inefficient auctions.
In 61.0\% of cases, winners win the auctions due to subsidization. In 48.0\% of cases, the builder with the highest true value shaded too much.
It is important to note that auctions may exhibit inefficiency for multiple reasons. For instance, the winner may subsidize her bid while simultaneously, the builder with the highest true value shades. Thus, the cumulative percentage exceeds 100\%.

It is important to note that inefficiency does not necessarily harm validators. In fact, validators may receive more revenues in inefficient scenarios involving subsidization. 

\parhead{MEV's implication on efficiency}
The ratio of efficient auctions in each tier of MEV is shown in~\autoref{fig:epi-mev-range}. We note that the trend of efficiency is less expected.
First, only 56.6\% of the auctions within the low MEV tier are efficient. This can be attributed to the generally low overall bid values, rendering the auction outcomes vulnerable to alterations from minor subsidies.
Second, while auctions tend to be more efficient as MEV increases, we still observe that 77.1\% of auctions in the high MEV tier are efficient, and about 86\% of inefficient cases result from bid shading, where the builders with the highest true values place their bids much lower than their true values.

\subsection{Measuring Block-building Capacities}
\label{sec:appendix-measuring-other-metrics}

\parhead{Coefficient of variation}
We use \(\mathsf{CV}(\vec{v})\to \mathbb{R}_{\geq 0}\) to denote the coefficient of variation (CV), a real value in \(\mathbb{R}_{\geq 0}\), for values $\vec{v}=(v_1,\dots,v_n)$. 
Specifically, suppose the true values of different builders in slot $s$ are \(\vec{tv}_s\). The CV of slot $s$ is computed as \(CV(\vec{tv}_s)\).

\begin{figure}[tbp]
    \centering
    \includegraphics[width=\columnwidth]{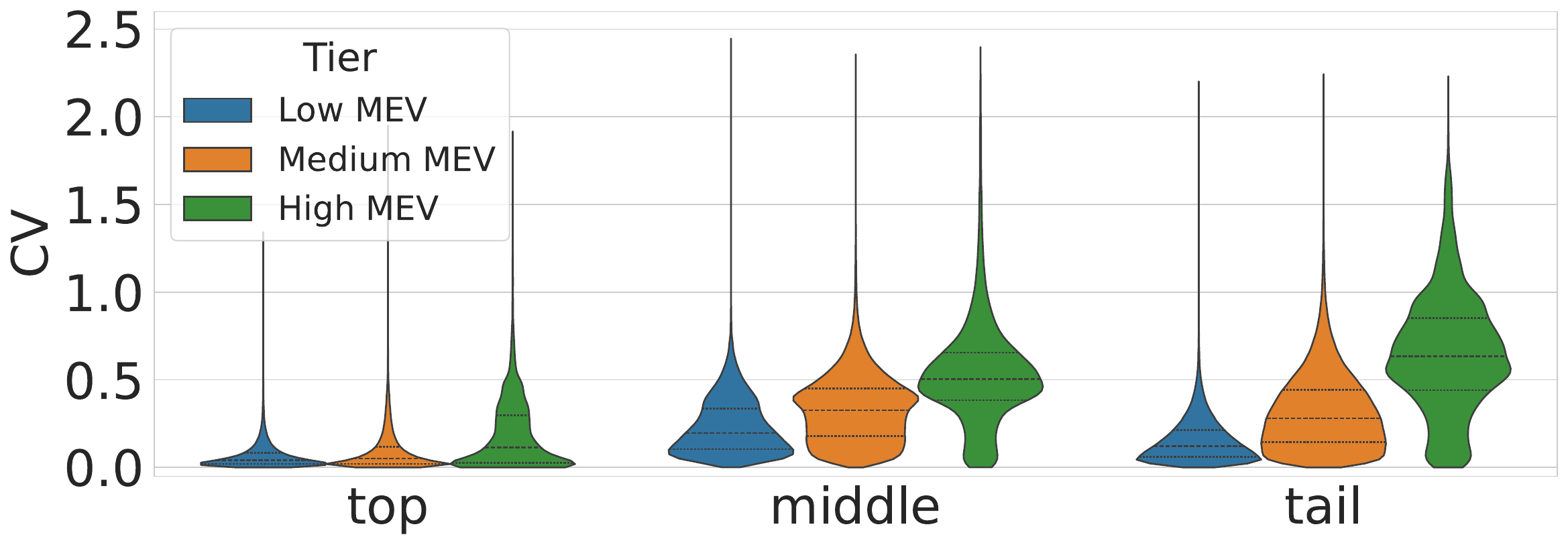}
    \caption{Violin graphs for the distribution of the CV across three MEV tiers, for top, middle, and tail builders. The CV measures the inequality of builders' block-build capacity.}
    \label{fig:topK-cv}
\end{figure}

As shown in~\autoref{fig:topK-cv}, we can make the same finding as~\autoref{fig:true-value-qcd}. It reveals that such inequality worsens as MEV increases. Within a given MEV tier, inequality is greatest among tail builders, followed by middle builders, and is least among top builders.

\subsection{Implications of Uncompetitive Auctions}

\begin{figure}[tbp]
    \centering
    \includegraphics[width=\columnwidth]{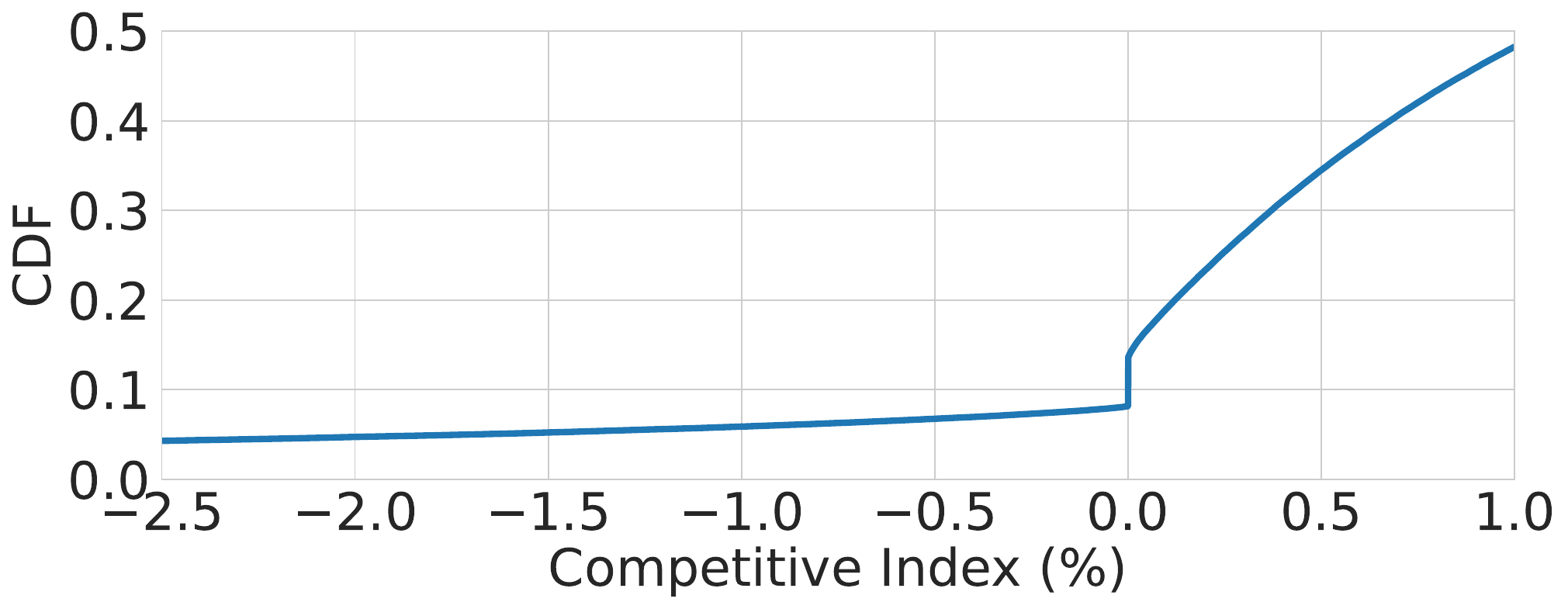}
    \caption{Distribution of the CIs of MEV-Boost auctions}
    \label{fig:ci-cdf}
\end{figure}

\autoref{fig:ci-cdf} shows the distribution of the competitive index (CI) of the MEV-Boost auctions in our study.

\section{Private Order Flows}
\label{appx:openness}

\begin{table*}[!htbp]
\centering
\scriptsize
\resizebox{\textwidth}{!}{
\begin{threeparttable}
\caption{Collections of the private order flow providers and their requirements.}
\label{tab:comparison-of-external-resources}
\begin{tabular}{p{1.4cm}|p{2.8cm}|p{12cm}}
\toprule
\textbf{Type} & \textbf{Provider} & \textbf{Requirements} \\
\midrule
Searcher & Individual searchers & Requirements are unclear. \\
\midrule
Telegram trading bot & Banana Gun~\cite{BananaGun2023}, Maestro~\cite{Maestro2023}, Unibot~\cite{UnibotX2023}  & The requirements are the same as the channels they use or unclear. \\
\midrule
\multirow{6}{*}{Channel} ~ & BloXroute~\cite{bloXrouteBackrunMe} & Builders are granted access to different tiers, each providing varying privileges and permissions, based on an assessment. \\
~ & Merkle~\cite{MerkleWebsite} & Builders must have a proven track record of accurately and honestly computing bundles, or they should consult with the Merkle team for consideration. \\
~ & MeowRPC~\cite{meowrpc} & The transactions are sent to the top builders which MeowRPC's team trusts.\\
~ & MEV Blocker~\cite{MEVBlocker} & Builders must maintain at least 1\% market share each week of the previous month. \\
~ & MEV-Share~\cite{FlashbotsMEVShare} & Builders must demonstrate a competitive market share for consideration. \\
~ & Transaction Boost~\cite{BlocknativeMEVProtection} & Consideration for partnership is based on the Blocknative team's evaluation of the builder's reputation. \\
\bottomrule
\end{tabular}
\begin{tablenotes}
    \item[$\star$] We do not claim that the list of the private order flow providers is complete since there is no ground truth.
\end{tablenotes}
\end{threeparttable}
}
\end{table*}
\subsection{Pivotal Private Order Flow Providers}
\label{appx:pivotal-providers}
We can make a few observations from~\autoref{fig:provider-of-crucial-transactions}. First, a particular searcher, \texttt{jaredfromsubway.eth}, has been gaining prominence since February 2023 and reached its peak significance in May 2023. During this period, the winners of more than 70\% of the MEV-Boost auctions would have failed had they not received order flow from \texttt{jaredfromsubway.eth}. Its influence has gradually decreased since then but still remains pivotal in 20\%-40\% of auctions since Nov 2023. %
Besides, the order flow from Telegram trading bots has become vital for builders with the rise of such bots. For instance, Banana Gun is pivotal in about 40\% of the MEV-Boost auctions in September 2023.

Second, we can observe a positive trend  from~\autoref{fig:provider-of-crucial-transactions} that MEV Share and MEV Blocker have been increasingly pivotal since May 2023. These two providers positively impact decentralization because they have relatively clear requirements for accessing their order flows (the requirements of different providers will be discussed next.) %

Third, we observe a notable downtrend in the pivotal level of MEV-Share and MEV Blocker starting from November 2023, accompanied by an uptrend in Maestro, a popular Telegram trading bot~\cite{Maestro2023}, beginning in December 2023. 
This is because Maestro used to be an important upstream of MEV-Share and MEV Blocker but stopped in November 2023, causing the pivotal levels of MEV-Share and MEV Blocker to drop significantly.

Although platforms like MEV-Share and MEV Blocker positively affect decentralization, the incentives of their upstream providers may not be aligned. For example, we observe that Maestro formed an integration with Beaver after ceasing to use these two platforms.

\subsection{Requirements of Providers}
\label{appx:subsidy}

Due to the lack of trustworthy fair exchange mechanisms, providers have to rely on an informal reputation system for self-protection, sending order flows only to builders who meet their requirements. To understand the access barrier caused by the trust crisis, we investigate different providers and their requirements for accessing order flows.
\autoref{tab:comparison-of-external-resources} summarizes the providers we identify and their requirements for accessing their private order flows.

\parhead{Subsidization.} New builders face a ``chicken and egg'' problem as they need access to private order flows to win auctions and gain market share, but the private order flow providers only serve builders with a decent market share.
In practice, builders pay out of pocket to overbid in MEV-Boost auctions, a practice known as block subsidization~\cite{relayscan2024}, representing an entry barrier for new builders.
We quantify the minimal cost for a new builder to meet the MEV Blocker's requirements (\appendixsubsecref{appx:quantify-barrier}), which reflects the costs of building a reputation. The minimal cost continues to increase and can be up to 18 ETH. This cost represents a lower bound since we do not have the corresponding numbers for other providers.
Note that although subsidization can help meet the requirements set by channels, it cannot ensure that searchers' trust concerns are addressed, allowing builders to receive their order flows.

\subsection{Minimal Cost for 1\% Market Share}
\label{appx:quantify-barrier}
\begin{figure}[htbp]
    \centering
    \includegraphics[width=\columnwidth]{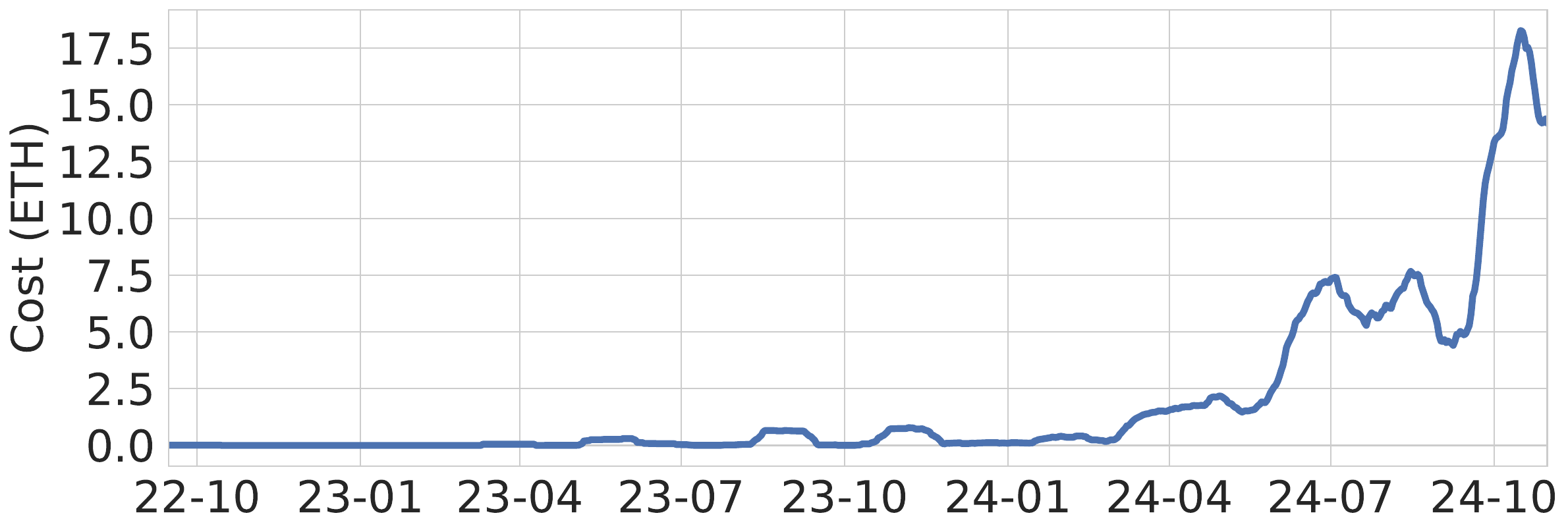}
    \caption{Subsidy required to establish a 1\% market share, as required by MEV Blocker. This quantifies the cost of accessing private order flows.}
    \label{fig:cost-of-mev-blocker}
\end{figure}

We compute the minimal subsidy needed to establish a 1\% market share for one month (which is the requirement of MEV Blocker~\cite{MEVBlocker}) by replaying historical auctions with an added new builder who can only access the public order flow.
We assume that all builders have access to public order flows but the new builder cannot access any of the private order flows. 
Furthermore, this new builder can win the auction if their bid value is the same as other builders. Other builders will not submit bids beyond their true value, defined as the total revenue of their blocks. Or they will not increase subsidies if they already subsidize.
For simplicity of computation, we assume that one month has 30 days.

Under our assumptions, a new builder must submit a bid at least equal to the auction winner's maximum--the higher of its bid value or true value.
The revenue from the new builder's block equals the value contributed by the public order flows to the winning bid. 
Thus, the subsidy needed for winning the auction is the difference between the revenue from public order flows and the required bid value. \done%
Then, as we know the subsidy for proposing one block, we can further calculate the minimum overall subsidy a new builder would need to cover to propose 72 blocks each day over one week (Ethereum produces 7200 blocks a day.)

The result is plotted in~\autoref{fig:cost-of-mev-blocker}. Initially, this cost was nearly zero before May 2023, but it has steadily increased, surpassing 18 ETH by October 2024.
It is important to recognize that the costs derived from this method only represent the lower bound. Predicting the minimum cost of winning a MEV-Boost auction is challenging, so builders need to pay more to ensure victory. Furthermore, builders may subsidize more blocks than required, as it is also difficult to precisely achieve a 1\% market share.

\end{document}